\documentclass[cmp,draft,numbook]{svjour}  
\usepackage[reqno,sumlimits]{amsmath}
\usepackage{amsfonts,amssymb,amsbsy}
\usepackage[final]{epsfig}
\usepackage{isolatin1}
\usepackage{submitextras}
\begin{document}
\title{The Absolute Continuity of the Integrated Density of States
for Magnetic Schr{\"o}dinger Operators with Certain Unbounded Random Potentials}
\titlerunning{Density of States for Random Schr{\"o}dinger Operators
  with Magnetic Fields}
\author{%
  Thomas Hupfer\inst{1},
  Hajo Leschke\inst{1},
  Peter M{\"u}ller\inst{2},        
  Simone Warzel\inst{1}}
\institute{%
  Institut f{\"u}r Theoretische Physik, 
  Universit{\"a}t 
  Erlangen-N{\"u}rnberg, Staudtstra{\ss}e 7, D--91058 
  Erlangen, Germany.
  \and
  Institut f{\"u}r Theoretische Physik,
  Georg-August-Universit{\"a}t, D-37073 G{\"o}ttingen,
  Germany.
  }            
\authorrunning{T. Hupfer, H. Leschke, P. M{\"u}ller, S. Warzel}
\dedication{{
    Dedicated to the memory of Kurt Broderix 
    (26 April 1962~~--~~12 May 2000)}}
\date{
  June 2001}
%
\maketitle
\begin{abstract}
The object of the present study is the integrated density of states 
of a quantum particle 
in multi-dimensional Euclidean space which is characterized
by a Schr{\"o}dinger operator with magnetic field and
a random potential which may be unbounded from above and below.
In case that the magnetic field is constant
and the random potential is ergodic and admits a so-called 
one-parameter decomposition,
we prove the absolute continuity of the integrated density
of states and provide explicit upper bounds on its derivative, the density 
of states.
This local Lipschitz continuity of the integrated density of states 
is derived by establishing a Wegner estimate for finite-volume Schr\"odinger operators 
which holds for rather general magnetic fields and different boundary conditions.
Examples of random potentials to which the results apply are 
certain alloy-type and Gaussian 
random potentials.
Besides we show a diamagnetic inequality for Schr{\"o}dinger 
operators with Neumann boundary conditions.
\end{abstract}

%
\tableofcontents
%
%
\section{Introduction} 
The integrated density of states is a quantity of primary interest in the 
theory \cite{Kir89,CaLa90,PaFi92} and 
application \cite{ShEf84,BoEn84,LiGr88,AnFoSt82,KuMeTi88}
of Schr{\"o}dinger operators 
for a particle in $d$-dimensional Euclidean space $\mathbbm{R}^d $ 
($ d = 1, 2, 3, \dots $)
subject to a random potential.
Its knowledge allows one to compute the free energy and hence all basic 
thermostatic quantities of the corresponding non-interacting 
many-particle system.
It also enters formulae for transport coefficients.
%

The main goal of the present paper is to prove the absolute continuity of the 
integrated density of states $ N $ 
for certain unbounded random potentials, 
thereby generalizing a result in \cite{FiHu97b} for zero magnetic field 
to the case of a constant magnetic field.
Examples of random potentials to which our result applies are 
certain alloy-type and Gaussian 
random potentials.
In particular, we consider the situation of 
two space dimensions and a 
perpendicular constant magnetic field 
where $ N $ 
is not absolutely continuous without random potential.

For the proof of absolute continuity of $ N $, we use the
abstract one-parameter spectral-averaging estimate of \cite{CoHi94} 
to derive what is called a Wegner estimate \cite{Weg81}. 
Such estimates provide upper bounds on the averaged number of eigenvalues 
of finite-volume random Schr{\"o}dinger operators in a given energy regime.
They play a major r{\^o}le in proofs of Anderson localization for
multi-dimensional random 
Schr{\"o}dinger operators \cite{CaLa90,PaFi92,CoHi94,FiLeMu00,Sto01}.  
In contrast to the Wegner estimates with magnetic fields which are available 
so far, we are neither restricted to the case 
of a constant magnetic field \cite{CoHi96,BaCoHi97b,Wan97}
nor to the existence of gaps in the spectrum of the magnetic
Schr{\"o}dinger operator
without random potential \cite{BaCoHi97a}.
In fact, the Wegner estimate in the present paper
holds for magnetic vector potentials whose components
 are locally square integrable.
%
Its proof involves techniques for (non-random) 
magnetic Neumann Schr{\"o}dinger operators among them
Dirichlet-Neumann bracketing and a diamagnetic inequality.
Appendix~\ref{AppNDI} provides the definition of these operators
and proofs of the latter techniques 
in greater generality than actually needed for the main body of
the present paper. 
%
\section{Random Schr{\"o}dinger Operators with Magnetic Fields}
\subsection{Basic notation}
As usual, let $\mathbbm{N} := \{1,2,3,\ldots\}$ denote the set of natural
numbers. 
Let $\mathbbm{R}$, respectively $\mathbbm{C}$,
denote the algebraic field of real, respectively complex numbers 
and let $ \mathbbm{Z}^d $ be the simple cubic lattice in $d $ dimensions, 
$d\in\mathbbm{N}$. An
open cube $\Lambda$ in $d$-dimensional Euclidean space $\mathbbm{R}^{d}$ is a translate of
the $d$-fold Cartesian product $ I\times\cdots\times I$ of
an open interval $I\subseteq\mathbbm{R}$.  
The open unit cube in $ \mathbbm{R}^d $ which is
centered at site $y \in \mathbbm{R}^d$ and whose edges are oriented parallel
to the co-ordinate axes is denoted by $ \Lambda(y) $.
The Euclidean norm of
$x\in\mathbbm{R}^{d}$ is  
$ |x|:=\big(\sum_{j=1}^{d}x_{j}^2\big)^{1/2} $.

The volume of a Borel subset $\Lambda\subseteq\mathbbm{R}^{d}$
with respect to the $d$-dimensional Lebesgue measure is $|\Lambda| :=
\int_{\Lambda}\d^{d}x = \int_{\mathbbm{R}^d}\!\d^{d}x \, \indfkt{\Lambda}(x)$,
where $ \indfkt{\Lambda} $ is the indicator function of $ \Lambda $.
In particular, if $ \Lambda $ is the strictly positive half-line,
$ \Theta := \indfkt{]\, 0 , \infty[} $ is the left-continuous Heaviside 
unit-step function.  
The Banach space ${\rm L}^{p}(\Lambda)$, $ p \in [ 1, \infty ] $, consists
of the Borel-measurable complex-valued functions 
$f: \Lambda\to\mathbbm{C}$ which are identified if their values differ only on a
set of Lebesgue measure zero and which obey $ \int_{\Lambda} \d^{d}x\, |f(x)|^{p} < \infty $ if $ p < \infty $
and $ \left\| f \right\|_\infty := \esssup_{x\in\Lambda} |f(x)| < \infty $ if $ p = \infty $.
We recall that ${\rm L}^{2}(\Lambda)$ is a
separable Hilbert space with scalar product 
$ \langle \cdot,\cdot \rangle $ given by   
$ \langle f, g \rangle = \int_{\Lambda}\d^{d}x\,  \overline{f(x)}\, g(x) $.
Here the overbar denotes complex conjugation. 
We write $f \in
{\rm L}^{p}_{\rm loc}(\mathbbm{R}^{d})$, if $ f \indfkt{\Lambda} \in {\rm L}^{p}(\mathbbm{R}^{d})$ for 
any bounded Borel set $\Lambda\subset \mathbbm{R}^{d}$.
Finally, $\mathcal{C}^{\infty}_{0}(\Lambda)$ is the vector
space of
functions $f: \Lambda\to\mathbbm{C} $ which are
arbitrarily often differentiable and have compact supports.

%
\subsection{Basic assumptions}
Let $ (\Omega, \mathcal{A}, \mathbbm{P}) $ be a complete probability space
and $ \mathbbm{E}\{ \cdot \} := \int_{\Omega} \! \mathbbm{P}(\d\omega)( \cdot ) $ be the expectation
induced by the probability measure $\mathbbm{P}$.
By a \emph{random potential} we mean a (scalar) random field
$ V:\, \Omega \times \mathbbm{R}^{d} \to \mathbbm{R} \, $,
$ (\omega ,x)\mapsto V^{(\omega )}(x) $ 
which is assumed to be jointly measurable with respect to the product of the 
sigma-algebra $ \mathcal{A} $ of event sets in $\Omega$ 
and the sigma-algebra $\mathcal{B}(\mathbbm{R}^d)$ 
of Borel sets in $ \mathbbm{R}^d $. 
We will always assume $ d \geq 2 $, because magnetic fields in one space 
dimension may be ``gauged away'' and are therefore of no physical relevance.
Furthermore, for $ d = 1 $ far more is known \cite{CaLa90,PaFi92} thanks to
methods which only work for one dimension.

We list four properties which $ V $ may have or not:
\begin{indentnummer*}
\item[\ass{F}] There exists some real $p \in ]1, \infty[$ with
  $p > 1$ if $ d = 2 $ and $ p \geq d/2 $ if $d \geq 3 $ such that
  for $\mathbbm{P}$-almost each $\omega \in \Omega$ 
  the realization $V^{(\omega)}: \, x \mapsto V^{(\omega)}(x) $  of $ V $ belongs to 
  ${\rm L}^p_{\rm loc}(\mathbbm{R}^d)$.
\item[\ass{S}] There exists some pair of reals 
  $ p_1 > p(d) $ and $ p_2 > p_1 d / \left[2 ( p_1 - p(d) ) \right]$ 
  such that 
  \begin{equation}\label{Eq:I}
    \sup_{y \in \mathbbm{Z}^d} \, \mathbbm{E} \Big\{ \big[
    \int_{\Lambda(y)} \!\!\! \d^d x \, |V(x)|^{\, p_1} \big]^{p_2/p_1} \Big\} 
    < \infty.
  \end{equation}
  Here $ p(d) $ is defined as follows:  
  $ p(d) := 2 $ if $d \leq 3 $, $ p(d) := d/2$ if $d \geq 5$ and $ p(4) > 2$, otherwise arbitrary.
\item[\ass{E}] $ V $ is $ \mathbbm{Z}^d $-ergodic or $ \mathbbm{R}^d $-ergodic.
\item[\ass{I}] The finiteness condition 
  \begin{equation}
    \sup_{y \in \mathbbm{Z}^d} \, \mathbbm{E} \big[
    \int_{\Lambda(y)} \!\!\!  \d^d x \, |V(x)|^{2\vartheta+1} \big] < \infty
  \end{equation}
  holds, where $\vartheta \in \mathbbm{N}$ is the smallest integer with $ \vartheta > d/4$.
\end{indentnummer*}
\begin{remark}
  \begin{nummer}
  \item
    Property \ass{E} requires the existence of a group 
    $ \mathcal{T}_x $, $x \in \mathbbm{Z}^d$  or 
    $  \mathbbm{R}^d $,
    of probability-preserving and ergodic transformations
    on $ \Omega $ such that $ V $ is 
    $ \mathbbm{Z}^d $- or $ \mathbbm{R}^d $-homogeneous in the sense that
    $ V^{(\mathcal{T}_x \omega)}(y) = V^{(\omega)}(y-x) $ 
    for all $ x \in \mathbbm{Z}^d $ or $ \mathbbm{R}^d $, 
    all $ y \in \mathbbm{R}^d $ and all $ \omega \in \Omega $. 
  \item
    Since property \ass{S} assures that the realization $V^{(\omega)}$ belongs to 
    ${\rm L}^{p(d)}_{\rm loc}(\mathbbm{R}^d)$ for $\mathbbm{P}$-almost each $\omega \in \Omega$, 
    property \ass{S} implies property \ass{F}.
    Property \ass{I} also implies property \ass{F}. 
  \end{nummer}
\end{remark}
We proceed by listing two properties either of which a random potential 
may additionally have or not and which
characterize two examples of random potentials, 
which we will consider in the present paper. 
\begin{indentnummer*}
\item[\ass{A}] $ V $ is an \emph{alloy-type random field}, that is, 
  a random field with realizations given by
  \begin{equation}
    V^{(\omega)}(x) = \sum_{j \in \mathbbm{Z}^d} \lambda_j^{(\omega)}
    u_0(x -j). 
  \end{equation}
  The \emph{coupling strengths} $\{ \lambda_j \} $ form a family
  of random variables which are $ \mathbbm{P} $-independent 
  and identically distributed according to the common
  probability measure $ \mathcal{B}(\mathbbm{R}) \ni I \mapsto \mathbbm{P}\{ \lambda_0 \in I \} $.
  Moreover, we suppose that the \emph{single-site potential}
    $ u_0: \mathbbm{R}^d \to \mathbbm{R}  $ satisfies the Birman-Solomyak condition
    ~$ \sum_{j \in \mathbbm{Z}^d} \big(\int_{\Lambda(j)} \! \d^d x \, |u_0(x)|^{p_1} \big)^{1/p_1} < \infty $
    with some real $ p_1 \geq 2 \vartheta + 1 $ and that 
    $ \mathbbm{E}\left( | \lambda_0 | ^{p_2} \right) < \infty $ for some real $ p_2 $ 
    satisfying $ p_2  \geq 2 \vartheta + 1 $ and $ p_2 > p_1 d /[2 (p_1 -p(d) )] $. 
  [The constants $ p(d) $ and $ \vartheta  $ are defined in properties~\ass{S} and \ass{I}.] 
\item[\ass{G}] $ V $ is a \emph{Gaussian random field} \cite{Adl81,Lif95}
  which is $ \mathbbm{R}^d $-homogeneous.
  It has  zero mean, $ \mathbbm{E}\left[\,V(0)\right] = 0 $, 
  and its covariance function
  $ x \mapsto C(x):= \mathbbm{E}\left[\, V(x) V(0) \right] $ 
  is continuous at the origin where it obeys 
  $ 0 < C(0) < \infty $.
\end{indentnummer*}
\begin{remark}
  \begin{nummer}
  \item
    Consider an \emph{alloy-type random potential} $ V $, that is,  
      a random potential with property~\ass{A}. 
    Then $ V $ has properties~\ass{E}, \ass{I}, \ass{S} and \ass{F}, see, for example \cite{HLMW01}. 
    \item
      Consider a random field with the Gaussian property~\ass{G}. 
      Then its covariance function $ C $ is bounded and 
      uniformly continuous on $ \mathbbm{R}^d $. 
      Consequently, \cite[Thm.\ 3.2.2]{Fer75} implies the existence of 
      a separable version $ V $ of this field which is jointly  measurable.
      Speaking about a \emph{Gaussian random potential},
      we tacitly assume that only this version will be dealt with.
      By the Bochner-Khintchine theorem \cite[Thm.~IX.9]{ReSi80} there is a one-to-one correspondence
      between finite positive (and even) 
      Borel measures on $ \mathbbm{R}^d $ and Gaussian random potentials. 
      An explicit calculation shows that a Gaussian random potential 
      enjoys properties~\ass{I}, \ass{S} and \ass{F}. 
      A simple sufficient criterion for
      the ergodicity property \ass{E} 
      is the mixing condition $\lim_{|x| \to \infty} C(x) = 0$.
    \end{nummer}
  \end{remark}
By a \emph{vector potential} we mean a (non-random) Borel-measurable vector field
$A : \mathbbm{R}^d \to  \mathbbm{R}^d$, $x \mapsto A(x)$ which we assume to possess
either the property
\begin{indentnummer*}
\item[\ass{B}] $\left| A \right|^2 $ belongs to ${\rm L}^1_{\rm loc}(\mathbbm{R}^d)$,
\end{indentnummer*}
or the property
\begin{indentnummer*}
\item[\ass{C}] $ A $ has continuous partial derivatives 
  which give rise to a magnetic field (tensor) with \emph{constant} components 
  given by $B_{jk}:=\partial_j A_k - \partial_k A_j$, where 
  $j$, $k \in \{1, \dots , d\}$. 
\end{indentnummer*}
\begin{remark}
  \begin{nummer}
  \item Property \ass{C} implies property \ass{B}.
  \item\label{RGauge}
    Given property \ass{C}, we may exploit the gauge freedom 
    to choose the vector potential in the \emph{symmetric gauge} 
    in which the components of $A$ are given by $A_k(x) = \sum_{j=1}^d x_j B_{jk}/2$, 
    where $k \in \{1, \dots , d\}$.
  \end{nummer}
\end{remark}
\subsection{Definition of the operators}
We are now prepared to precisely define 
magnetic Schr{\"o}dinger operators with random
potentials on the Hilbert spaces $ {\rm L}^2(\Lambda)$ and $ {\rm L}^2(\mathbbm{R}^d) $. 
The finite-volume case is treated in 
\begin{proposition}\label{PropDefHL}
  Let $\Lambda \subset \mathbbm{R}^d$ be a bounded 
  open cube, $A $ be a
  vector potential with the property \ass{B} and  $V$ be a random 
  potential with the property \ass{F}. 
  Then 
  \begin{indentnummer*}
  \item
    the sesquilinear form
    \begin{equation}
      h_{\Lambda, {\rm N}}^{A, 0} (\varphi , \psi) := \frac{1}{2} \sum_{j=1}^{d} 
      \left\langle ( {\i} \nabla + A)_j \, \varphi \, , 
        \, ( {\i} \nabla + A)_j \, \psi \right\rangle,
    \end{equation}
    with $\varphi$, $\psi$ in the form domain
    $\mathcal{Q}\bigl(h_{\Lambda,{\rm N}}^{A, 0}\bigr) := 
    \left\{ \phi \in {\rm L}^2(\Lambda) : 
      ({\i} \nabla + A) \, \phi \in \right.$\hspace{0pt}$\left.({\rm L}^2(\Lambda))^d \right\}$ 
    and $ \nabla - {\i} A $ denoting the gauge-covariant gradient in the sense of 
    distributions on $ \mathcal{C}_0^\infty ( \Lambda ) $, uniquely defines
    a self-adjoint positive operator on ${\rm L}^2(\Lambda)$, which we denote by $ H_{\Lambda,{\rm N}}(A,0)$. 
    The closure $ h_{\Lambda , {\rm D}}^{A,0} $ of the restriction of 
    $ h_{\Lambda , {\rm N}}^{A,0} $ 
    to the domain $  \mathcal{C}_0^{\infty }\left( \Lambda \right) $ 
    uniquely defines another self-adjoint positive operator
    on ${\rm L}^2(\Lambda)$,
    which we denote by $H_{\Lambda,{\rm D}}(A,0)$.
  \item 
    The two operators
    \begin{equation}
      H_{\Lambda,{\rm X}}(A,V^{(\omega)}) := H_{\Lambda,{\rm X}}(A,0) + V^{(\omega)}, 
      \qquad {\rm X=D}\;\; {\rm or} \;\; {\rm X=N}, 
    \end{equation}
    are self-adjoint and bounded below on ${\rm L}^2(\Lambda)$ as form sums
    for all $\omega$ in some subset $\Omega_{\sf F} \in \mathcal{A}$ of $\Omega $
    with full probability, in symbols,
    $\mathbbm{P}(\Omega_{\sf F}) = 1$.
  \item 
    The mapping 
    $H_{\Lambda,X}(A, V): \Omega_{\sf F}  \ni \omega \mapsto H_{\Lambda,{\rm X}}(A,V^{(\omega)})$
    is measurable. We call it the 
    \emph{finite-volume magnetic Schr{\"o}dinger operator with random potential $V$} 
    and Dirichlet or Neumann boundary condition if $\rm X = D$ or 
    $\rm X = N$, respectively.
  \item
    The spectrum of $H_{\Lambda,\rm X}(A,V^{(\omega)})$ is purely discrete for all 
    $ \omega \in \Omega_{\sf F}$.
  \item
    The (random)  
    \emph{finite-volume density-of-states measure}, defined by the trace 
    \begin{equation}\label{Def:nu}
      \nu_{\Lambda,{\rm X}}^{(\omega)}(I):=
      \Tr\Big[ \indfkt{I}\big(H_{\Lambda,{\rm X}}(A ,V^{(\omega)})\big) \Big],
    \end{equation}
    is a positive Borel measure on the real line $\mathbbm{R}$ for all $\omega \in \Omega_{\sf F}$.  
    Here $\indfkt{I}\big(H_{\Lambda,{\rm X}}(A ,V^{(\omega)})\big)$ is 
    the spectral projection operator of $H_{\Lambda,{\rm X}}(A ,V^{(\omega)})$ 
    associated with the energy regime $I\in \mathcal{B}(\mathbbm{R})$. 
    Moreover, the
    (unbounded left-continuous) distribution function
    \begin{equation}\label{eq:finite-volumeIDOS}
      N_{\Lambda , \rm X}^{(\omega)}(E) :=  
      \nu_{\Lambda,{\rm X}}^{(\omega)}\big(\left] - \infty, E \right[ \big) 
      = \Tr\Big[ \Theta\big( E - H_{\Lambda,{\rm X}}(A ,V^{(\omega)}) \big) \Big] < \infty
    \end{equation}
    of $ \nu_{\Lambda,{\rm X}}^{(\omega)} $,
    called the \emph{finite-volume integrated density of states},
    is finite for all energies $ E \in \mathbbm{R} $. 
  \end{indentnummer*}
\end{proposition}
%
%
\begin{proof}
  The proofs of assertions~(i), (ii) and (iv) 
  are contained in Appendix~\ref{AppNDI} because \ass{B} and \ass{F} imply
  (\ref{eq:ass:av+}) and (\ref{eq:ass:v-}).
  Assertion (iii) is a consequence of considerations in 
  \cite{KirMar82a}, see also
  Sect.~V.1 in \cite{CaLa90}, 
  and of a straightforward generalization to non-zero 
  vector potentials.
  Assertion (v) follows from (ii) and (iv).
  \qed
\end{proof}
\begin{remark}
  Counting multiplicity, $\nu_{\Lambda,{\rm X}}^{(\omega)}(I)$ is just the number of 
  eigenvalues of the operator $H_{\Lambda,\rm X}(A ,V^{(\omega)})$ 
  in the Borel set $ I \subseteq \mathbbm{R} $. Since this number is almost-surely finite if $ I $ is bounded, 
  the mapping $ \nu_{\Lambda,{\rm X}} : \omega \mapsto  \nu^{(\omega)}_{\Lambda,{\rm X}} $ 
  is a random Borel measure.
\end{remark}
The precise definition of the infinite-volume magnetic Schr{\"o}dinger operator  
on ${\rm L}^2(\mathbbm{R}^d)$ and a compilation of its basic properties are given in
\begin{proposition}\label{Prop:DefH}
  Assume that the vector potential
  $A$ and the random potential $V$ enjoy the properties \ass{C} and \ass{S}.
  Then
  \begin{indentnummer*}
  \item
    the 
    operator $\mathcal{C}_0^\infty(\mathbbm{R}^d) \ni \psi \mapsto 
    \frac{1}{2} \sum_{j = 1}^d ({\i} \partial_j + A_j)^2 \, \psi + V^{(\omega)}\psi$ is essentially self-adjoint
    for all $\omega$ in some subset $\Omega_{\sf S} \in \mathcal{A}$ 
    of $ \Omega $ with full probability,
    $\mathbbm{P}(\Omega_{\sf S}) = 1$. Its self-adjoint closure 
    on ${\rm L}^2(\mathbbm{R}^d)$
    is denoted by $H(A,V^{(\omega)})$. 
  \item
    The mapping
    $H(A, V): \Omega_{\sf S}  \ni \omega \mapsto H(A,V^{(\omega)})$
    is measurable. 
    We call it the 
    \emph{infinite-volume magnetic Schr{\"o}dinger operator with random potential $V$}.
  \end{indentnummer*}
\end{proposition}
\begin{proof}
  For assertion~(i) see \cite[Prop.~2.3]{FiLeMu00}, which generalizes 
  \cite[Prop.~V.3.2]{CaLa90} to the case of continuously differentiable
      vector potentials $ A \neq 0$. Note that the assumption
      of a vanishing divergence, $ \sum_{j = 1}^d \partial_j A_j = 0 $, in 
      \cite[Prop.~2.3]{FiLeMu00} is not needed in the argument.
  Assertion~(ii) is a straightforward generalization of \cite[Prop.~V.3.1]{CaLa90} to
      continuously differentiable $ A \neq 0 $,
      see also \cite[Prop.~2 on p.~288]{Kir89}. \qed 
\end{proof}
\begin{remark}
  For alternative or weaker criteria instead of \ass{S} guaranteeing 
  the almost-sure self-adjointness
  of $ H(0, V) $, see \cite[Thm.~5.8]{PaFi92} or \cite[Thm.~1 on p.~299]{Kir89}.
\end{remark}
If $ A $ has the property~\ass{C}, the infinite-volume magnetic 
Schr{\"o}dinger operator without scalar potential, $H(A,0)$, 
is unitarily invariant under so-called \emph{magnetic translations} \cite{Zak64}.
The latter form a family of unitary operators 
$\{T_x\}_{x \in \mathbbm{R}^d}$ on ${\rm L}^2(\mathbbm{R}^d)$
defined by
\begin{equation}
  \left(T_x \psi\right) (y) := \e^{{\i} \Phi_x(y)} \psi(y - x ) , 
  \qquad \qquad \psi \in {\rm L}^2(\mathbbm{R}^d),
\end{equation}
where ~$\Phi_x  (y) :=  \int_{\mathcal{K}(x,y)} \d r \cdot \left( A(r) - A(r-x) \right) $~
is an integral along some smooth curve $\mathcal{K}$ 
with initial point $x \in \mathbbm{R}^d$ and terminal point $y \in \mathbbm{R}^d$. 
Since $A$ and its $x$-translate $A(\,\cdot -x)$ give rise to the same magnetic field 
and $\mathbbm{R}^d$ is simply connected, 
the integral $\Phi_x(y)$ is actually independent of $\mathcal{K}$.
\begin{remark}
  \begin{nummer}
  \item For the vector potential 
    in the symmetric gauge (see Remark~\ref{RGauge})
    one has $\Phi_x(y) = \sum_{j,k=1}^d x_j B_{jk} (y_k - x_k)/2$.
  \item For a discussion in the case of more general configuration spaces 
    and magnetic fields,
    see for example \cite{MohRai94}.
  \item In the situation of Prop.~\ref{Prop:DefH}
    and if the random potential $V$ has property \ass{E}, we have
    \begin{equation}
      T_x \, H(A, V^{(\omega)}) \, T_x^\dagger =  H(A, V^{(\mathcal{T}_x \omega)})
    \end{equation}
    for all $ \omega \in \Omega_{\sf S}$ and all $ x \in \mathbbm{Z}^d $ or $ x \in \mathbbm{R}^d $, 
    depending on whether $ V $
    is $ \mathbbm{Z}^d $- or $ \mathbbm{R}^d $-ergodic. Hence, following standard arguments, $ H(A, V) $ is an
    \emph{ergodic operator} and its spectral components are non-random,
    see \cite[Thm. 2.1]{Uek94}. Moreover, 
    the discrete spectrum of $ H(A, V^{(\omega)}) $ is empty for $\mathbbm{P}$-almost 
    all $\omega\in \Omega$, 
    see \cite{Kir89,CaLa90,Uek94}.
  \end{nummer}
\end{remark}
%
%
\subsection{The integrated density of states}
The quantity of main interest in the present paper is 
the integrated density of states
and its corresponding measure, called the density-of-states measure.
The next theorem, which we recall from \cite{HLMW01}, deals with its definition and its 
representation as an infinite-volume 
limit of the suitably scaled finite-volume counterparts~(\ref{eq:finite-volumeIDOS}). 
\begin{proposition}\label{Thm:IDOS}
  Let
  $\indfkt{\Lambda(0)}$ denote the multiplication operator
  associated with the indicator function of the unit cube $ \Lambda(0) $.
  Assume that the potentials $A$ and $V$ have the
  properties \ass{C}, \ass{S}, \ass{I} and \ass{E}.  
  Then the \emph{(infinite-volume) integrated density of states} 
  \begin{equation}\label{Eq:DefIDOS}
    N(E) :=  \mathbbm{E}\Big\{
    \Tr\big[ \indfkt{\Lambda(0)} \, \Theta\big(E - H(A,V) \big) 
      \, \indfkt{\Lambda(0)} \big]\Big\} < \infty
  \end{equation}
  is well defined for all energies $ E \in \mathbbm{R} $ in terms of the (spatially localized) 
  spectral family of the infinite-volume operator $ H(A, V) $.
  It is 
  the (unbounded left-continuous) distribution function 
  of some positive Borel measure $ \nu $ 
  on the real line $ \mathbbm{R}$.
  Moreover,  let $\Lambda \subset \mathbbm{R}^d$ stand for bounded open cubes centered at 
  the origin. Then there is a set $ \Omega_0 \in \mathcal{A} $ 
  of full probability, $ \mathbbm{P}(\Omega_0) = 1 $, such that the limit relation  
  \begin{equation}\label{Eq:IDOSconv}
    N(E) =  \lim_{\Lambda \uparrow \mathbbm{R}^d} \, 
    \frac{N_{\Lambda,{\rm X}}^{(\omega)}(E)}{\left| \Lambda \right|}
  \end{equation}
  holds for both boundary conditions ${\rm X} = {\rm D}$
  and ${\rm X} = {\rm N}$, all $ \omega \in \Omega_0 $ and 
  all $ E \in \mathbbm{R} $ 
  except for the (at most countably many) discontinuity points of $ N $.
\end{proposition}
\begin{proof}
  See \cite{HLMW01}. \qed
\end{proof}
\begin{remark}
  \begin{nummer}
  \item\label{Re:IDOS} A proof of the \emph{existence} of the 
    integrated density of states $ N $ under slightly different hypotheses was
    outlined in \cite{Mat93}. It uses functional-analytic arguments
    first presented in \cite{KirMar82} for the case $ A = 0 $.
    A different approach to the existence of the
    density-of-states measure $ \nu $ for $ A \neq 0 $,
    using Feynman-Kac(-It{\^o}) functional-integral representations of
    Schr{\"o}dinger semigroups \cite{Sim79,BrHuLe00}, 
    can be found in \cite{Uek94,BrHuLe93}.
    The latter approach dates back to \cite{Pas71,Nak77} for the case $ A = 0 $.
    To our knowledge, it works straightforwardly in the case $ A \neq 0 $ 
    for $ \rm X = \rm D $ only. 
    For $ A \neq 0 $ the \emph{independence} of the infinite-volume limit in
    (\ref{Eq:IDOSconv}) of the boundary
    condition $ \rm X $ (previously claimed without proof in \cite{Mat93}) follows 
    from \cite{Nak00} if  
    the random potential $ V $ is bounded and from \cite{DoIwMi01} if $ V $ is
    bounded from below. 
    So the main new point about Prop.~\ref{Thm:IDOS} is that it also applies to 
    a wide class of $ V $ unbounded from below.
    Even for $ A = 0 $, Prop.~\ref{Thm:IDOS} is partially new 
    in that the corresponding result \cite[Thm.~5.20]{PaFi92} only shows vague convergence of the underlying measures, 
    see the next remark. 
  \item
      An immediate corollary of Prop.~\ref{Thm:IDOS}
      is the 
      vague convergence \cite[Def.~30.1]{Bau92} of 
      the \emph{spatial eigenvalue concentrations}
      $ \left| \Lambda \right|^{-1} \, \nu_{\Lambda, \rm X}^{(\omega)} $
      in the infinite-volume limit $ \Lambda \uparrow \mathbbm{R}^d $
      to the non-random positive Borel 
      measure $ \nu $ uniquely corresponding to the integrated density of states 
      (\ref{Eq:DefIDOS}) in the sense that
      $ N(E) = \nu ( ] - \infty , E [) $ for all $ E \in \mathbbm{R} $,
      that is,
      \begin{equation}\label{Eq:DOSvag}
        \nu = \lim_{\Lambda \uparrow \mathbbm{R}^d} \, 
        \frac{ \nu^{(\omega)}_{\Lambda,{\rm X}}}{\left| \Lambda \right|} \qquad \quad\mbox{(vaguely)}
      \end{equation}
      for  both ${\rm X} = {\rm D}$
      and ${\rm X} = {\rm N}$ and $ \mathbbm{P} $-almost all $ \omega \in \Omega $.
\end{nummer}
\end{remark}
%
%

%
One may relate properties of the 
density-of-states measure $ \nu $
to simple spectral properties of the infinite-volume
magnetic Schr{\"o}dinger operator. 
Examples are the support of $\nu$ and the 
location of the almost-sure spectrum of $ H(A, V^{(\omega)}) $ or  
the absence of a point component in the Lebesgue decomposition of $\nu $ and 
the absence of ``immobile eigenvalues'' of $H(A, V^{(\omega)}) $.
This is the
content of
\begin{corollary}\label{Thm:Stet}
  Under the assumptions of Prop.~\ref{Thm:IDOS} 
  and letting $ I \in \mathcal{B}(\mathbbm{R}) $,
  the following
  equivalence holds:
  ~$  \nu(I) = 0 $~ if and only if ~$\indfkt{I}\big(H(A, V^{(\omega)})\big) = 0 $~
   for $ \mathbbm{P}$-almost all $\omega\in \Omega$.
  This immediately implies:
  \begin{indentnummer*}
    \item
      ~$ \supp \nu = \spec H(A, V^{(\omega)}) $ for $\mathbbm{P}$-almost all $\omega \in \Omega$.
      {\rm[}Here $ \spec H(A, V^{(\omega)}) $ denotes the  spectrum of  $  H(A, V^{(\omega)}) $ and 
      ~$\supp \nu := \{ E \in \mathbbm{R} :  \nu(]E - \varepsilon, E + \varepsilon[) > 0 \;\;
      \mbox{\rm for all} \; \varepsilon > 0 \} $ is 
      the topological support of $ \nu $.{\rm]} 
    \item
      ~$0 = \nu(\{E\}) \; \big[ = \lim_{\varepsilon \downarrow 0} \left[ N( E + \varepsilon) - N(E) \right] \big]$ 
      ~if and
      only if $E\in \mathbbm{R}$ is not an eigenvalue of $H(A,V^{(\omega)})$
      for $\mathbbm{P}$-almost all $\omega \in \Omega$.
  \end{indentnummer*}
\end{corollary}
\begin{proof} 
  See \cite{HLMW01}. \qed
\end{proof}
The equivalence~(ii) of the above corollary is a continuum analogue 
of \cite[Prop.~1.1]{CrSi83}, see also \cite[Thm.~3.3]{PaFi92}.  In the
one-dimensional case \cite{Pas80} and the multi-dimensional lattice 
case \cite{DelSou84}, the equivalence has been exploited to show  for $ A = 0 $ the
(global) continuity of the integrated density of states $ N $ 
under practically no further assumptions on the random potential beyond 
those ensuring the existence of $ N $. The proof of such a statement in the
multi-dimensional continuum case is considered an important open
problem \cite{Sim00}. 
For $ A \neq 0 $ one certainly needs additional assumptions
as \cite{DoMa99} illustrates, 
see Remark~\ref{Rem:DMP99} below.  
Under the additional assumptions of Corollary~\ref{Cor:ExDOS} below,
we will show that the integrated density of states 
is not only continuous, but even absolutely continuous in the case of a constant magnetic field of arbitrary strength.

%
\section{Existence of the Density of States for Certain Random Potentials}\label{Ch:3}
In this section we provide
conditions under which the integrated density of states $ N $ 
(or, equivalently, its measure $ \nu $) is absolutely continuous
with respect to the Lebesgue measure.
As a by-product, we get rather explicit upper bounds on the resulting 
Lebesgue density $ \d N(E) / \d E = \nu( \d E ) / \d E $, called the 
density of states.
Results of this genre date back to \cite{Weg81} and go nowadays under the
name Wegner estimates.
\subsection{A Wegner estimate}
The main aim of this subsection is to extend the 
Wegner estimate in \cite{FiHu97b} 
to the case with magnetic fields.
For this purpose we recall from there
\begin{definition}\label{Def:decomposition}
  A random potential $ V: \Omega \times \mathbbm{R}^d \to \mathbbm{R} $ admits a  
  ~$ (U, \lambda , u, \varrho ) $-\emph{decomposition} if there exists a random 
  potential $ U: \Omega \times \mathbbm{R}^d \to \mathbbm{R}\, $, a random 
  variable $ \lambda : \Omega \to \mathbbm{R} $ and a real-valued
  $ u \in {\rm L}^{\infty}_{\rm loc}(\mathbbm{R}^d) $ such that
  \begin{indentnummer*}
  \item $ V^{(\omega )} = U^{(\omega )} + \lambda^{(\omega )} u $ 
    ~~for $\mathbbm{P}$-almost all $ \omega \in \Omega $,
  \item the conditional probability distribution of $ \lambda $ relative 
    to the sub-sigma-algebra generated by the family of random variables 
    $ \{ U(x) \}_{x\in\mathbbm{R}^d} $ has a 
    jointly measurable density 
    $ \varrho : \Omega \times \mathbbm{R} \to [0, \infty[ $ 
    with respect to the Lebesgue 
    measure on $ \mathbbm{R}\, $. 
  \end{indentnummer*}
\end{definition}
The condition $ u \in {\rm L}^{\infty}_{\rm loc}(\mathbbm{R}^d) $ was 
missed out in \cite[Def.~2]{FiHu97b}.
We now state the following generalization of \cite[Thm.~2]{FiHu97b} 
which in its turn
relies on a result in \cite{CoHi94}.
\begin{theorem}\label{ThmWegner}
  Let $ \Lambda \subset \mathbbm{R}^d $ be a bounded open cube.
  Let $ \Lambda = \left(\; \overline{ 
      \bigcup_{j=1}^{J} \Lambda _{j}} \;\right)^{\rm int} $ be decomposed into the 
  interior of the closure of finitely many, $ J \in \mathbbm{N} $,
  pairwise disjoint bounded open 
  cubes $ \Lambda _{j} \subset \mathbbm{R}^d $.
  Let the potentials $A$ and $V$  be supplied with the properties \ass{B} and \ass{F},
  respectively.
  Assume that for each $ j \in \{ 1, \dots , J \} $
  the random potential $ V $  
  admits a ~$ (U_{j}, \lambda _{j}, u_{j},
  \varrho _{j}) $-decomposition
  subject to the following three conditions: 
  there exist five strictly positive constants $ v_1 , v_2, \beta ,R ,Z > 0$ 
  such that 
  for all $  j \in \{ 1, \dots , J \} $,
  \begin{nummer}
    \item
      $  v_1 \indfkt{\Lambda _{j}}(x) \leq u_{j}(x) $
      and $  u_{j}(x) \indfkt{\Lambda _{j}}(x) \leq v_2 $
      for Lebesgue-almost all $ x\in \mathbbm{R}^d $,
    \item
       $ \displaystyle \esssup_{\xi \in\mathbbm{R}} 
       \Big( \varrho _{j}^{(\omega )}(\xi )\,
       \max\{ \e^{-\beta v_1 \xi } ,
       \e^{-\beta v_2 \xi }\} \Big)\leq R $ 
       ~~for $\mathbbm{P}$-almost all $ \omega \in \Omega $,
    \item
      $ \displaystyle \mathbbm{E}\left\{ \Tr \left[ 
          \e^{-\beta H_{\Lambda_j , {\rm N}}(A, U_j)} 
      \right] \right\} \leq \left| \Lambda_j \right| \, Z  $.
  \end{nummer}
  Then
  the averaged number of eigenvalues of the finite-volume operator  
  $ H_{\Lambda , {\rm X}}(A, V) $ in any non-empty energy regime 
  $I \in \mathcal{B}(\mathbbm{R}) $ of
  finite Lebesgue measure $ | I | $ is bounded from above according to 
  \begin{equation}\label{Eq:Wegner}
    \mathbbm{E}\left[ \, \nu_{\Lambda,X}(I) \right] \leq 
    \left| \Lambda \right| \, 
    |I|  \, \,  
    \frac{R Z}{v_1} \, \e^{\beta \sup I}
  \end{equation}
 for both boundary conditions $X$. 
 {\rm [}Here $\sup I$ denotes the least upper bound of $I \subset \mathbbm{R}$.{\rm]}
\end{theorem}
\begin{remark}\label{Rem:Tscheby}
    The (Chebyshev-Markov) inequality 
    $ \indfkt{[1, \infty[}(\left| \xi \right| ) \leq \left| \xi \right| $ implies
    \begin{equation}\label{Eq:Tscheby}
      \mathbbm{P}\Big\{ 
      I \, \cap \, \spec H_{\Lambda , {\rm X}}(A, V) 
      \neq \emptyset \Big\} 
      = \mathbbm{E}\Big[ 
      \indfkt{[1, \infty[}\big( \nu_{\Lambda, {\rm X}}(I) \big) \Big]
      \leq
      \mathbbm{E} \Big[ \nu_{\Lambda, {\rm X}}(I)  \Big].
    \end{equation}
    Therefore the Wegner estimate (\ref{Eq:Wegner}) 
    in particular bounds the probability of finding
    at least one eigenvalue of $  H_{\Lambda , {\rm X}}(A, V) $ 
    in a given energy regime $ I \in \mathcal{B}(\mathbbm{R}) $.
    Such bounds are a key ingredient of proofs of 
    Anderson localization for multi-dimensional random Schr{\"o}dinger operators,
    see \cite{CaLa90,PaFi92,CoHi94,FiLeMu00,Sto01} and references therein.
\end{remark}
\begin{proof}[of Theorem~\ref{ThmWegner}]
  Since we follow exactly the strategy of the proof 
  of \cite[Thm.~2]{FiHu97b}, we only 
  remark that the two main steps in this proof remain valid 
  in the presence of a vector potential $A$. 
  The first step, used in inequality (27) of \cite{FiHu97b}, concerns
  the lowering of the eigenvalues of the operator $H_{\Lambda, \rm X}(A, V^{(\omega)})$
  by so-called Dirichlet-Neumann
  bracketing in case $X=D$ and by the (subsequent) insertion of
  interfaces in $ \Lambda $  
  with the requirement of Neumann boundary conditions. 
  For $A\neq 0$, supplied with property \ass{B}, 
  the validity of these two techniques is established in Appendix~\ref{AppNDI}.
  The second step is an application of a
  spectral-averaging estimate of \cite{CoHi94}, 
  which is re-phrased as Lemma~\ref{LSM} below.
  Since there the operator $ L $ is only required to be
  self-adjoint and does not enter the r.h.s. of (\ref{eq:LSM}), 
  it makes no difference if $ L $ is taken as 
  $H_{\Lambda, \rm X}(0,U_j)$ (as is done in \cite{FiHu97b}) 
  or as $H_{\Lambda, \rm X}(A,U_j)$ for each $j \in \{1, \dots, J\}$.
\qed
\end{proof}
  An essential tool in the preceding proof is the 
  (simple extension of the) abstract one-parameter 
  \emph{spectral-averaging estimate} of \cite{CoHi94};
  in this context see also \cite{CoHiMo96}.
\begin{lemma}\label{LSM}
  Let $K$, $L$ and $M$ be three self-adjoint operators 
  acting on a Hilbert space $\mathcal{H}$
  with $K$ and $M$ bounded such that
  ~$ \kappa := \inf_{K \varphi \neq 0} \,  
    \langle \varphi \, , \, M  \, \varphi \rangle / 
    \langle \varphi \, , \, K^2 \,  \varphi \rangle > 0 $ is strictly positive.
  Moreover, let $ g \in {\rm L}^\infty(\mathbbm{R})$. 
  Then the inequality
  \begin{equation}\label{eq:LSM}
    \int_{\mathbbm{R}} \! \d \xi \, \, \left| g(\xi) \right| \, \langle \psi \, , \, 
    K \, \indfkt{I}(L + \xi M) \, K \, \psi \rangle 
    \leq |I| \, \frac{\|g\|_\infty}{\kappa} \, \langle \psi , \psi \rangle
  \end{equation}
  holds for all $\psi \in\mathcal{H}$ and all $I \in \mathcal{B}(\mathbbm{R})$. 
\end{lemma}
\begin{proof}
Since the assumption $ \kappa > 0 $ implies the operator inequalities $ 0 \leq \kappa \, K^2 \leq M $,
the lemma is proven as Cor.~4.2 in \cite{CoHi94} for any positive bounded 
function $g$ with compact support.
It extends to positive bounded functions 
with arbitrary supports by a monotone-convergence argument. 
\qed 
\end{proof}
%
%
\subsection{Upper bounds on the density of states}
If the fraction $R Z /v_1 $ on the r.h.s of the Wegner estimate 
(\ref{Eq:Wegner}) is independent of 
$\Lambda$ for sufficiently large 
$ \left| \Lambda \right| $,
this estimate enables one to prove the absolute continuity of the 
infinite-volume density-of-states measure with
a magnetic field.
\begin{corollary}\label{Cor:ExDOS}
  Let $A$ and $V$ have the
  properties \ass{C}, \ass{S}, \ass{I} and \ass{E}. 
  Suppose furthermore:
  \begin{indentnummer*}
    \item there exists a sequence $ ( \Lambda ) $ of bounded open 
      cubes $ \Lambda \subset \mathbbm{R}^d $ 
      with $ \Lambda \uparrow \mathbbm{R}^d $ such that infinitely many
      of them admit a decomposition
      ~$\Lambda = \left(\; \overline{ \bigcup_{j=1}^{J} \Lambda _{j}}
        \;\right)^{\rm int} $ into a finite number $J$ (depending on
      $\Lambda$) of pairwise disjoint open cubes $\Lambda_1, \dots , \Lambda_J $.
    \item $V$ obeys the assumptions of
      Theorem~\ref{ThmWegner} for every such decomposition with constants 
      $\beta$, $v_1$, $R$, $Z > 0$,  
      all of them not depending on $ \Lambda $.
  \end{indentnummer*}
  Then the density-of-states measure $ \nu $ 
  is absolutely continuous with respect to the Lebesgue
  measure. Moreover, its Lebesgue density $ w $, called the \emph{density of
    states}, is locally bounded according to
  \begin{equation}\label{Eq:ExDOS}
    w(E) := \frac{\nu(\d E)}{\d E} = 
    \frac{\d N(E)}{\d E} \leq \frac{R Z}{v_1} \, \e^{\beta E} =: W(E)
  \end{equation}
  for Lebesgue-almost all energies $E \in \mathbbm{R}$.
\end{corollary}
\begin{proof}
  Let $ I \subset \mathbbm{R} $ be bounded and open. 
  Then (\ref{Eq:DOSvag}) together with \cite[Satz~30.2]{Bau92} 
  implies that 
  $ \nu(I) \leq  \liminf_{\Lambda \uparrow \mathbbm{R}^d}  
  |\Lambda|^{-1}  \nu^{(\omega)}_{\Lambda, {\rm X}}(I) $ 
  for $ \mathbbm{P} $-almost all $ \omega \in \Omega $.
  Therefore, by the non-randomness of the density-of-states measure $\nu $
  and Fatou's lemma we have 
  \begin{equation}
    \nu(I) \leq \liminf_{\Lambda \uparrow \mathbbm{R}^d} \,
    \frac{\mathbbm{E}\left[ \nu_{\Lambda, {\rm X}}(I) \right]}{|\Lambda|}
    \leq |I|  \, \,  
    \frac{R Z}{v_1} \, \e^{\beta \sup I}.
  \end{equation}  
  Here we used (\ref{Eq:Wegner}) and the assumption that the 
  constants involved there do not depend on $ \Lambda $. Now the Rad\'on-Nikod\'ym theorem yields the
  claimed absolute continuity of $ \nu $.
  \qed
\end{proof}

%
\section{Examples Illustrating the Results of Section~\ref{Ch:3}}
Assumption~(iii) of Theorem~\ref{ThmWegner} may be checked in various ways.
For example, by the diamagnetic inequality (\ref{eq:diazusum}) of Appendix~\ref{AppNDI}
for Neumann partition functions
one sees that a possible choice
of $ Z $ in (\ref{Eq:Wegner}) is 
\begin{equation}\label{eq:Z1}
  Z_1 := \max_{ 1\leq j\leq J} \,\left\{ |\Lambda_j|^{-1} \mathbbm{E}\left[ \Tr \left( 
        \e^{-\beta H_{\Lambda_j , {\rm N}}(0, U_j)} 
      \right) \right] \right\}.
\end{equation}
This yields an 
upper bound on $ \mathbbm{E}\left[ \, \nu_{\Lambda,X}(I) \right] $ in (\ref{Eq:Wegner}) which is 
\emph{independent} 
of the magnetic field and, in particular, coincides
with the one in \cite[Thm.~2]{FiHu97b}.
Rather weak conditions on the random potential $ U_j $ assuring the finiteness of the expectation value 
in (\ref{eq:Z1}) can be found in \cite{DrKi86}.

Another choice of $ Z $ results from 
applying the following averaged Golden-Thompson inequality.
\begin{lemma}\label{LemmaZS}
  Let $\Lambda\subset\mathbbm{R}^d$ be a bounded open cube and assume that 
  $A$ and $V$ enjoy properties \ass{B} and \ass{F}. 
  Then the
  \emph{averaged partition function} of $  H_{\Lambda,{\rm X}}(A,V) $
  is bounded for all $\beta > 0$ according to
  \begin{equation}\label{eq:ZS}
     \mathbbm{E} \left\{ \Tr\left[  \e^{- \beta \, H_{\Lambda,{\rm X}}(A,V)}\right] \right\}
     \leq \Tr\left[ \e^{ - \beta \, H_{\Lambda, {\rm X}}(A,0)} \right] \,\, 
     \esssup_{x \in \Lambda} \left\{\mathbbm{E}\left[ \e^{-\beta \, V(x)} \right]\right\},
  \end{equation}
  provided that the essential supremum on the r.h.s. is finite. 
\end{lemma}
\begin{proof}
  We proceed as in the proof of \cite[Thm.~3.4(ii)]{KirMar82} and define
  $V^{(\omega)}_n(x) := \max\{-n, V^{(\omega)}(x)\}$ for $n \in \mathbbm{N}$ and $ \omega \in \Omega_{\sf F} $.  
  The Golden-Thompson inequality \cite{ReSi78} yields
  \begin{equation}\label{eq:GT}
    \Tr\left[  \e^{- \beta \, H_{\Lambda,{\rm X}}(A,V^{(\omega)}_n)}\right]
    \leq  \Tr\left[ \e^{- \beta  H_{\Lambda,{\rm X}}(A,0)} \, \e^{- \beta \, V^{(\omega)}_n} \right]. 
  \end{equation}
  We then evaluate the trace on the r.h.s. in an orthonormal eigenbasis
  of $H_{\Lambda,{\rm X}}(A,0)$. 
  Using Fubini's theorem, the probabilistic expectation 
  of the quantum-mechanical
  expectation of ~$\exp(- \beta V_n)$~ with respect to a normalized eigenfunction 
  of $H_{\Lambda,{\rm X}}(A,0)$ is estimated by 
  ~$\esssup_{x \in \Lambda} \left\{\mathbbm{E}\left[ \exp(-\beta \, V_n(x)) \right]\right\}$, 
  which is smaller than the second factor on the r.h.s. 
  of (\ref{eq:ZS}) since $V \leq V_n$.
  The proof is completed by noting that the l.h.s. 
  of (\ref{eq:GT}) converges for $n \to \infty$
  to the trace on the
  l.h.s. of (\ref{eq:ZS}) by monotone convergence 
  of forms \cite[Thm. S.16]{ReSi80},
  similar to the proof of \cite[Prop.~2.1(e)]{KirMar82}. 
\qed
\end{proof}
Using (\ref{eq:ZS}) one gets
\begin{equation}\label{Eq:Z2}
  Z_2 := \max_{ 1\leq j\leq J} \,\left\{ |\Lambda_j|^{-1} \Tr\left[ 
      \e^{- \beta  H_{\Lambda_j, \rm N}(A,0)} \right] \,
    \esssup_{x \in \Lambda_j} \mathbbm{E}\left[ \e^{-\beta U_j(x)} \right] 
  \right\}
\end{equation}
as another choice for $Z$ in (\ref{Eq:Wegner}).
By (\ref{eq:diazusum})
one may further estimate the magnetic Neumann partition function in (\ref{Eq:Z2}) according to 
\begin{equation}\label{eq:zudoof}
  \Tr\left[ \e^{ - \beta \, H_{\Lambda,{\rm N}}(A,0)} \right] 
  \leq \Tr\left[ \e^{ - \beta \, H_{\Lambda,{\rm N}}(0,0)} \right] 
  \leq |\Lambda| \big( |\Lambda|^{-1/d} + (2\pi \beta)^{-1/2}\big)^d.
\end{equation}
The second inequality follows from 
the explicitly known \cite[p.~266]{ReSi78} spectrum of
$ H_{\Lambda,{\rm N}}(0,0) $.
Applying (\ref{eq:zudoof}) to (\ref{Eq:Z2}) one 
weakens $ Z_2 $ to a rather explicit choice 
of $Z$ in (\ref{Eq:Wegner}) given by
\begin{equation}\label{Def:Z3}
  Z_3 := \max_{ 1\leq j\leq J}\,\left\{ 
    \left( |\Lambda_j|^{-1/d} + (2 \pi \beta)^{-1/2}\right)^d \,
    \esssup_{x \in \Lambda_j} \mathbbm{E}\left[ \e^{-\beta U_j(x)} \right] 
  \right\}.
\end{equation}
\subsection{Alloy-type random potentials}
The existence of a $(U,\lambda, u, \varrho)$-decomposition of $V$ as required
in Theorem~\ref{ThmWegner} is immediate for alloy-type random potentials whose
coupling strengths are distributed according to a Borel probability measure
on the real line  
with a bounded Lebesgue density. 
To illustrate the essentials of Theorem~\ref{ThmWegner} we first consider
the case of positive potentials.
\begin{corollary}\label{Cor:AT}
  Let $A$ and $V$ have the properties~\ass{B} and \ass{A}. Assume that 
  $ u_0 \in {\rm L}^\infty_{\rm loc}(\mathbbm{R}^d) $ and that 
  the probability distribution of $ \lambda_0 $ has a 
  Lebesgue density $ g \in  {\rm L}^\infty(\mathbbm{R})$
  with support in the positive half-line~$ \left[ 0, \infty \right[$.
  Furthermore, suppose that there 
  exist two strictly positive constants $v_1$, $v_2 > 0$ such that
  \begin{equation}\label{As:Traeger}
    v_1 \indfkt{\Lambda(0)}(x) \leq u_0(x) 
    \quad \mbox{and} \quad
    u_0(x) \indfkt{\Lambda(0)}(x) \leq v_2
  \end{equation}
  for Lebesgue-almost all $ x\in \mathbbm{R}^d $.
  Then for each bounded open 
  cube of the form 
  \begin{equation}\label{eq:Kubus}
    \Lambda = \Big(\; \overline{ 
      \bigcup_{j \in \Lambda \cap \mathbbm{Z}^d} \Lambda(j) } \;\Big)^{\rm int},
  \end{equation}
  one has
  \begin{equation}\label{Eq:AT}
    \mathbbm{E} \left[ \, \nu_{\Lambda, \rm X}(I) \right] \leq \left| \Lambda \right| \,
    \left| I \right| W_{\sf A}(\, \sup I)
  \end{equation}
  for both $ \rm X = \rm D $ and $ \rm X = \rm N $ 
  and all $I \in \mathcal{B}(\mathbbm{R})$.
  Here $ W_{\sf A} $ is the function 
  \begin{equation}
    \mathbbm{R} \ni E \mapsto  W_{\sf A}(E) :=  \left( 1 + (2 \pi \beta)^{-1/2} \right)^d \,
      \frac{\left\| g \right\|_\infty}{v_1} \,
      \e^{\beta E}
  \end{equation}
  with $ \beta  \in ]\, 0, \infty [$ serving as a variational parameter.
\end{corollary}
\begin{proof}
  For each 
  $j \in  \Lambda \cap \mathbbm{Z}^d$, 
  the choice ~$ u_j(x) := u_0(x -j)$ and
  ~$ U^{(\omega)}_j(x) := V^{(\omega)}(x) - \lambda^{(\omega)}_j u_j(x)$
  yields a ~$ (U_j, \lambda_j, u_j, g) $-decomposition of $V $ 
  in the sense
  of Definition~\ref{Def:decomposition}. 
  It remains to verify the three assumptions of Theorem~\ref{ThmWegner}.
  Assumption~(i) is guaranteed by (\ref{As:Traeger}). 
  Assumption~(ii) is fulfilled with $ R = \left\| g \right\|_\infty $. 
  To verify assumption~(iii), we
  make use of (\ref{Def:Z3}) and observe that $U^{(\omega)}_j \geq 0$.
  \qed
\end{proof}
\begin{remark}
  \begin{nummer}
  \item\label{Rem:Correlated}
    The estimates in the proof of Corollary~\ref{Cor:AT},
    when specializing the fraction $ R Z /v_1 $ of 
    Theorem~\ref{ThmWegner} to $ W_{\sf A} $,
    were unnecessarily rough for the sake of simplicity. 
    In specific examples the upper bound $ W_{\sf A} $ may be improved.
    Moreover, more general alloy-type random potentials are also covered
    by Theorem~\ref{ThmWegner}. 
    In particular, the random potential may be unbounded from below, 
    see the next corollary. 
    Furthermore, one may allow for
    correlated coupling strengths $\{\lambda_{j}\}$ as long
    as the relevant conditional probabilities 
    have bounded Lebesgue densities.  
  \item
    Apart from the existence of a bounded Lebesgue density for the coupling strength $ \lambda_0 $ 
    one further restrictive assumption of Corollary~\ref{Cor:AT} 
    is the fact that the single-site potential $u_0$ must possess a definite
    sign.
    The latter may be slightly weakened such that one may treat
    certain $ u_0 $ taking on values of both signs 
    by choosing a more complicated decomposition different from the natural one used in the proof of 
    Corollary~\ref{Cor:AT}. 
    This basically corresponds to the linear-transformation technique introduced in \cite{Ves00} 
    which turns certain given alloy-type random potentials into ones with positive single-site potentials 
    and correlated coupling strengths,
    see the previous Remark~\ref{Rem:Correlated}.
    In any case, the fact that $ u_0 $ must possess a sufficiently large support
    is believed to be important for the absolute continuity
    of the integrated 
    density of states in the presence of a magnetic field,
    see Remark~\ref{Rem:DMP99}. 
  \item
    We only know of \cite{CoHi96,BaCoHi97a,BaCoHi97b,Wan97} 
    where Wegner estimates for 
    magnetic Schr{\"o}dinger operators with alloy-type random potentials 
    have been derived.\footnote{See, however, note added in proof.} 
    The Wegner estimate of \cite{BaCoHi97a} is proven for energies in
    pre-supposed gaps of the spectrum of $ H(A,0) $.
    The other three works consider the case of two space dimensions and a 
    perpendicular constant magnetic field, see Subsect.~\ref{Sec:KonstB},
    especially Remark~\ref{Rem:CoHi} and \ref{Rem:Wa}.
  \end{nummer}
\end{remark}    
We close this subsection by considering the example 
of an unbounded below alloy-type random potential 
with exponentially decaying probability density for its (independent)
coupling strengths.
This example is marginal in the sense that any such density 
has to  fall off at minus infinity at least as fast as exponentially 
in order to ensure the applicability of Theorem~\ref{ThmWegner}.
\begin{corollary}\label{Ex:Exp}
  Let $ A $ and $ V $ have the properties~\ass{B} and \ass{A}. 
  Assume a Laplace distribution for $ \lambda_0 $, that is
  \begin{equation}
     \mathbbm{P} \big\{\lambda_0 \in I \big\} =  \frac{1}{ 2 \alpha }
     \, \int_{I}\! \d \xi \,\, 
     \e^{ - | \xi |/ \alpha }, 
     \qquad I \in \mathcal{B}(\mathbbm{R}),
  \end{equation}
  with some $\alpha > 0$. 
  Furthermore, suppose that $ u_0 \in {\rm L}^\infty(\mathbbm{R}^d) $ 
  and that {\rm (\ref{As:Traeger})} holds with some 
  $v_1$, $v_2 > 0$ and let 
  \begin{equation}
    K_\beta :=  - \essinf_{x \in \Lambda(0)} \sum_{j \in \mathbbm{Z}^d} 
    \ln \left\{ 1 - [ \beta \alpha u_0( x -j )]^2 \right\} < \infty
  \end{equation}
  be finite for some $\beta \in ]\, 0, (\alpha \| u_0 \|_\infty )^{-1}[$. 
  Finally, let $ \Lambda $ be of the form {\rm (\ref{eq:Kubus})}. 
  Then {\rm (\ref{Eq:AT})} holds where
  $ W_{\sf A} $ may be taken as the function
  \begin{equation}
     E \mapsto W_{\sf A}(E) := \left( 1 + (2 \pi \beta)^{-1/2} \right)^d \,
      \frac{ 1 - (\beta \alpha v_1)^2}{2 \alpha v_1} 
       \e^{\beta E + K_\beta}
  \end{equation}
  with $ \beta \in \left\{ \beta' \in \, ]\, 0, (\alpha \| u_0 \|_\infty )^{-1}[ \, : \,  K_{\beta'} < \infty \right\} $ 
  serving as a variational parameter.
\end{corollary}
\begin{proof}
  The proof is analogous to that of Corollary~\ref{Cor:AT}.
  To verify the assumptions
  of Theorem~\ref{ThmWegner} we note that 
  assumption~(i) is guaranteed by (\ref{As:Traeger}). Assumption~(ii)
  is fulfilled with $ R = (2 \alpha)^{-1} $ if $\beta \in ]\, 0, (\alpha v_2)^{-1}]$. 
  As for assumption~(iii), we
  make use of (\ref{Def:Z3}) and explicitly compute the involved 
  expectation if $\beta \in ]\, 0, (\alpha \| u_0 \|_\infty )^{-1}[$.
  \qed
\end{proof}
\subsection{Gaussian random potentials}
As another application of Theorem~\ref{ThmWegner} 
we note that the Wegner estimate 
derived previously  \cite[Thm.~1]{FiHu97b} 
for certain Gaussian random potentials and the case without magnetic field 
remains valid in the present setting.
The reason for this is the fact that 
\emph{every} Wegner estimate stemming from \cite[Thm.~2]{FiHu97b} is also
one in the presence of a magnetic field thanks to the diamagnetic inequality.
\begin{corollary}\label{Cor:Gauss}
  Let $A$ and $V$ have the properties~\ass{B} and \ass{G}. 
  Moreover, assume that there exist a finite signed
  Borel measure $\mu$ on $\mathbbm{R}^d$, which is normalized in the sense that
  $ \int_{\mathbbm{R}^d}  \mu(\d^dx) \,  
  \int_{\mathbbm{R}^d}  \mu(\d^dy) \,  C(x-y) = C(0) $, an open subset 
  $ \Gamma \subset \mathbbm{R}^d$ with volume $ \big| \Gamma \big| > 0 $ and 
  a constant $ \gamma > 0 $ such that the covariance function $ C $ of $ V $ obeys
  \begin{equation}\label{Eq:Defu}
     \gamma \,  
    \indfkt{\Gamma}(x) \leq \left( C(0) \right)^{-1} \, 
    \int_{\mathbbm{R}^d} \! \mu(\d^dy) \, C(x-y) =: \left( C(0) \right)^{-1/2}
    u(x)
  \end{equation}
  for all $x \in \mathbbm{R}^d $. Then for each $ \ell > 0 $, for which there exists a 
  bounded open cube $ \Lambda^{(\ell)} \subseteq \Gamma $ with edges of length $ \ell $ 
  parallel to the co-ordinate axes, and each bounded open cube 
  $ \Lambda \subset \mathbbm{R}^d $ satisfying the matching condition  
  $ \left|  \Lambda \right|^{1/d} / \ell \in \mathbbm{N} $, one has
  \begin{equation}
    \mathbbm{E} \left[ \nu_{\Lambda, {\rm X}}(I) \right]
    \leq \left| \Lambda \right| \, \left| I \right| \, W_{\sf G}( \, \sup I )
  \end{equation}
  for both $ \rm X = \rm D $ and $ \rm X = \rm N $ and all 
  $ I \in \mathcal{B}(\mathbbm{R}) $. Here $ W_{\sf{G}} $ is the function 
  \begin{equation}\label{eq:Gaussbound}
    E \mapsto W_{\sf{G}}(E) 
    :=  \left( 2 \ell^{-1} + (2 \pi \beta)^{-1/2} \right)^d \;
    \frac{  \exp\left\{\beta E + \beta^2 C_\ell/2\right\}}{\sqrt{2 \pi C(0)}\, b_{\ell}}
  \end{equation}
  where we introduced the constants
  $C_{\ell} := C(0) \left( 1 + B_{\ell}^2 - b_{\ell}^2 \right)$, 
  $B_{\ell} := \left( C(0) \right)^{-1/2} \, \sup_{x \in \Lambda^{(\ell)} }\, u(x) $
  and 
  $ b_{\ell} := \left( C(0) \right)^{-1/2} \, \inf_{x \in \Lambda^{(\ell)} }\, u(x) \geq \gamma $.
  Finally, $ \beta \in ]\, 0 , \infty [$ serves, besides $ \ell $, as a second variational parameter.
\end{corollary}
\begin{proof}
  The key input is the fact
  that every Gaussian random potential $ V $ admits a  
  $(U,\lambda, u ,\varrho)$-decomposition in the sense of 
  Definition~\ref{Def:decomposition}. More precisely, 
  $ \lambda^{(\omega)} :=  \left( C(0) \right)^{-1/2} \, 
  \int_{\mathbbm{R}^d} \mu(\d^dx) V^{(\omega)}(x) $ is a standard Gaussian
  random variable with Lebesgue density 
  $ \varrho(\xi) := \left( 2 \pi \right)^{-1/2} \, 
  \exp\left( - \xi^2 / 2 \right) $.
  This random variable and the 
  Gaussian random field $ U^{(\omega)}(x) :=  V^{(\omega)}(x) - 
  \lambda^{(\omega)} u(x) $, where
  $ u $ is defined in (\ref{Eq:Defu}), are stochastically independent.
  For details see the proof of \cite[Thm.~1]{FiHu97b}.
  To obtain the specific form $ W_{\sf G} $, 
  which is independent of the magnetic field, we used (\ref{Def:Z3}).
  \qed
\end{proof}
\begin{remark}
\begin{nummer}
  \item
  Without loss of generality, every measure $ \mu $ yielding (\ref{Eq:Defu}) 
  may be normalized in the sense of the assumption in the 
  above corollary. 
  The measure $ \mu $ allows one to apply Corollary~\ref{Cor:Gauss}
  to Gaussian random potentials with certain covariance functions taking on also 
  negative values. Examples are given in \cite{FiHu97b,HuLeWa00}.  
  \item
  If $C(x) \geq 0$ for all $x \in \mathbbm{R}^d$, 
  we may choose $ \mu $ equal 
  to Dirac's point measure at the origin. Due to the continuity of $ C $
  and since $ C(0) > 0 $, condition~(\ref{Eq:Defu}) is then fulfilled with
  some sufficiently small cube $ \Gamma $ containing the origin and
  $ \gamma = \inf_{x \in \Gamma} C(x)/C(0) $.
  Under 
  stronger conditions on the vector potential $ A $ the Wegner estimate for this
  case has been stated in \cite[Prop.~2.14]{FiLeMu00} where
  it serves as one input for a proof of Anderson localization by certain
  Gaussian random potentials, see Remark~\ref{Rem:Tscheby}.
  \item
  Choosing $\ell = | E |^{-1/4} $
  and
  $ \beta =  \left( 2 C_{\ell} \right)^{-1} \left( \sqrt{ E^2 + 2 d \, C_{\ell} } -E \right) $
  we obtain the
  following leading 
  low- and high-energy behaviour:
  \begin{equation}\label{Eq:GaussAsym}
    \lim_{E \to - \infty} \, \frac{\ln W_{\sf{G}}(E)}{E^2} = - \frac{1}{2 C(0)},
    \qquad \quad
    \lim_{E \to \infty} \, \frac{W_{\sf{G}}(E)}{E^{d/2}} =  
    \frac{\left( e/(\pi d) \right)^{d/2}}{\sqrt{2 \pi }\, u(0)}.
  \end{equation}
  Since $ W_{\sf{G}} $ provides an upper bound on the density of states
  (see Corollary~\ref{Cor:ExDOS}), its 
  low-energy behaviour is optimal in the sense
  that it coincides with that of the 
  derivative of the known low-energy behaviour of the 
  integrated density of states \cite{Mat93,Uek94,BrHuLe93}.
  This is not true for the high-energy behaviour. 
  It is known \cite{Mat93,Uek94} that 
  the high-energy growth of the 
  integrated density of states is neither affected
  by the random potential 
  nor by the magnetic field 
  and proportional to $ E^{d/2}$ for $E \to \infty$
  in analogy to
  Weyl's celebrated asymptotics for the free particle \cite{Wey12}. 
  Note that the
  constant on the r.h.s. of the second equation in (\ref{Eq:GaussAsym})
  is smaller than the one
  given by \cite[Eq.~(14)]{FiHu97b}.
\end{nummer}
\end{remark}
%
%
%
\subsection{Two space dimensions: random Landau Hamiltonians}\label{Sec:KonstB}
In this subsection we consider the special case of two space dimensions
and a  perpendicular constant magnetic field of strength $B:=B_{12}>0$.
Accordingly, the vector potential
in the symmetric gauge is given by 
\begin{equation}
  A(x)= \frac{B}{2}\,\binom{-x_2}{x_1}, \qquad x = \binom{x_1}{x_2} \in \mathbbm{R}^2.
\end{equation}
This case has received considerable attention during 
the last three decades \cite{AnFoSt82,KuMeTi88}
in the physics of low-dimensional electronic structures.

The magnetic Schr{\"o}dinger operator on ${\rm L}^2(\mathbbm{R}^2)$ modelling the 
non-relativistic motion  of
a particle with unit charge on the Euclidean plane $ \mathbbm{R}^2 $  
under the influence of this magnetic field 
is the \emph{Landau Hamiltonian}. 
Its spectral resolution dates back to Fock \cite{Foc28} 
and Landau \cite{Lan30} and
is given by the strong-limit relation
\begin{equation}\label{Eq:LandauHam}
H(A,0) = \frac{B}{2} \, \sum_{\ll=0}^{\infty} \left(2 \ll + 1\right) P_\ll.
\end{equation}
The energy eigenvalue $(\ll + 1/2) B$ is called 
the $\ll^{\rm th}$ \emph{Landau level} and the corresponding 
orthogonal eigenprojection $P_\ll$ is an
integral operator with continuous complex-valued kernel
\begin{equation}
  P_\ll(x,y) := \frac{B}{2\pi} \exp\left[{\rm i} \frac{B}{2}  (x_2 y_1 - x_1 y_2) 
    - \frac{B}{4} |x-y|^2\right] 
  {\rm L}_\ll\left(\frac{B}{2}|x-y|^2\right),
\end{equation}
given in terms of the $ \ll^{\rm th} $  Laguerre polynomial
$\xi \mapsto {\rm L}_\ll(\xi):=\frac{1}{\ll !} {\rm e}^\xi  \frac{\rm d^\ll}{{\rm d} \xi^\ll} 
\left( \xi^{\ll} {\rm e}^{-\xi}\right)$,
$\xi \geq 0$, \cite[Sect.~8.97]{GrRy}. 
The diagonal $P_\ll(x,x)= B/(2\pi)$ is naturally interpreted as the degeneracy
per area of the $ \ll^{\rm th} $ Landau level.

\begin{figure}[hbt]
\begingroup\makeatletter%
\gdef\SetFigFont#1#2#3#4#5{%
  \reset@font
  \fontsize{8}{#2pt}
  \fontfamily{#3}\fontseries{#4}\fontshape{#5}%
  \selectfont}%
\endgroup%
\begin{center}
\begin{picture}(0,0)%
\includegraphics{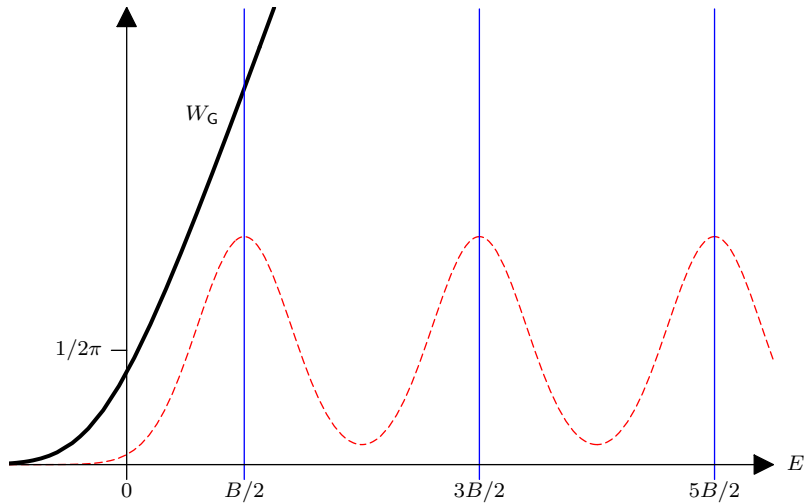}%
\end{picture}%
\setlength{\unitlength}{4144sp}%
\begingroup\makeatletter\ifx\SetFigFont\undefined%
\gdef\SetFigFont#1#2#3#4#5{%
  \reset@font\fontsize{#1}{#2pt}%
  \fontfamily{#3}\fontseries{#4}\fontshape{#5}%
  \selectfont}%
\fi\endgroup%
\begin{picture}(4680,2937)(429,-2536)
\put(1160,-2536){\makebox(0,0)[b]{\smash{\SetFigFont{8}{9.6}{\familydefault}{\mddefault}{\updefault}
$0$
}}}
\put(1864,-2536){\makebox(0,0)[b]{\smash{\SetFigFont{8}{9.6}{\familydefault}{\mddefault}{\updefault}
$B/2$
}}}
\put(3271,-2536){\makebox(0,0)[b]{\smash{\SetFigFont{8}{9.6}{\familydefault}{\mddefault}{\updefault}
$3B/2$
}}}
\put(4677,-2536){\makebox(0,0)[b]{\smash{\SetFigFont{8}{9.6}{\familydefault}{\mddefault}{\updefault}
$5B/2$
}}}
\put(5109,-2371){\makebox(0,0)[lb]{\smash{\SetFigFont{8}{9.6}{\familydefault}{\mddefault}{\updefault}
$E$
}}}
\put(1511,-281){\makebox(0,0)[lb]{\smash{\SetFigFont{12}{14.4}{\familydefault}{\mddefault}{\updefault}
$W_\mathsf{G}$
}}}
\put(1016,-1686){\makebox(0,0)[rb]{\smash{\SetFigFont{8}{9.6}{\familydefault}{\mddefault}{\updefault}
$1/2\pi$
}}}
\end{picture}
\caption{Plot of the upper bound $ W_{\sf G}(E) $ on $ w(E) $
  as a function of the energy $E$.
  Here $ w $ is 
  the density of states of the Landau Hamiltonian 
  with a Gaussian random potential with Gaussian covariance function. 
  The dashed line is a plot of the graph of an approximation to $ w $.
  The exact $ w $ is unknown. 
  Vertical lines indicate the delta peaks which reflect the 
  non-existence of the density of states 
  without random potential $ V $.
  The step function $ \Theta(E) / 2\pi $ (not shown) 
  is the free density of states characterized by 
  $ B = 0 $ and $ V = 0 $. (See text).
  }\label{bild}
\end{center}
\end{figure}

Using definition~(\ref{Eq:DefIDOS}) with $ V = 0 $, 
the integrated density of states of the Landau Hamiltonian (\ref{Eq:LandauHam})
turns out to be the well-known ``staircase'' function
\begin{equation}\label{eq:LandauIDOS}
  N(E) = \frac{B}{2 \pi} \, \sum_{\ll=0}^\infty \, \Theta\Big(E - \Big(\ll + \frac{1}{2}\Big)B\Big),
  \qquad V=0,
\end{equation}
which is obviously not absolutely continuous 
with respect to the Lebesgue measure.
For the derivation of (\ref{eq:LandauIDOS}) 
one may apply \cite[Thm.~VI.23]{ReSi80} 
because the operator $ P_\ll \, \indfkt{\Lambda(0)} $ is Hilbert-Schmidt, more precisely  
$ \Tr [ \indfkt{\Lambda(0)} P_\ll \indfkt{\Lambda(0)} ] = B / (2 \pi ) < \infty $. 
Alternatively one may compute \cite[App.~B]{Nak00} the infinite-area limit
$ \lim_{\Lambda \uparrow \mathbbm{R}^2} \, \left| \Lambda \right|^{-1} \, 
\Tr [ \Theta(E - H_{\Lambda, \rm X}(A,0)) ] $
for some boundary condition $ \rm X $. 
The result coincides with (\ref{eq:LandauIDOS}) by Prop.~\ref{Thm:IDOS}.
Informally, the density of states associated with (\ref{eq:LandauIDOS}) 
is a series of Dirac delta functions supported at the Landau levels.
The corresponding infinities are indicated 
by vertical lines in Fig.~\ref{bild} 
and together form what might be called a ``Dirac half-comb''.  
By adding a random potential $V$ to (\ref{Eq:LandauHam}),
the delta peaks are expected to be smeared out. 
In fact, under the assumptions of 
Corollary~\ref{Cor:ExDOS} they are smeared out completely in the sense that
the density of states $ w $ of the arising
\emph{random Landau Hamiltonian} $H(A,V) = H(A,0) + V$ 
is shown there to be locally bounded.\\

For example, in the presence of a Gaussian random potential with the 
Gaussian covariance function $ C(x) = C(0) \exp\big\{- |x|^2 /(2 \tau^2) \big\} > 0$, 
$\tau > 0 $,  Fig.~\ref{bild} contains the graph of the upper bound 
$ W_{\sf G} $ on $ w $ given in (\ref{eq:Gaussbound}) 
after (numerically) minimizing  
with respect to $\beta$, $ \ell $ and a certain one-parameter 
subclass of possible decompositions of $ V $.
Here we picked 
a (small) disorder parameter, $ C(0) = (B/5)^2 $, and 
a (large) correlation length, $ \tau = 100 B^{-1/2} $.
We recall that the function $  W_{\sf G} $ is independent 
of $ B $ due to our application of the diamagnetic inequality, 
but nevertheless
provides an upper bound on $ w $ for all $ B \geq 0 $.
Therefore  $  W_{\sf G}(E) $ is a rather rough estimate of $ w(E) $
already for energies $ E < B/2  $ and, in particular,
starts increasing significantly at too low energies.
Nevertheless, the upper bound shows that 
the density of states $ w $ has no 
infinities for arbitrarily weak
disorder, that is, for arbitrarily small $ C(0) > 0 $.
In fact, in the above situation
we believe the graph of $w$ to look similar 
to the dashed line in Fig.~\ref{bild}. \\

We conclude this subsection with several remarks:
\begin{remark}
  \begin{nummer}
  \item 
    Unfortunately, our upper bound $W$ in (\ref{Eq:ExDOS}) is never sharp enough to reflect 
    the expected ``magneto-oscillations'' of $w$. 
    Instead, by construction $ W $ is always increasing.
  \item\label{Rem:DMP99}
    The assumptions of Corollary~\ref{Cor:ExDOS} 
    guarantee in particular that there occurs no point component in the  Lebesgue decomposition
    of the density-of-states measure $ \nu $. 
    Using Corollary~\ref{Thm:Stet}, this implies that any given energy $ E \in \mathbbm{R} $, 
    in particular any Landau-level energy, 
    is $\mathbbm{P}$-almost surely no eigenvalue 
    under these assumptions.
    This stands in contrast to a certain situation with random 
    point impurities, 
    in which case the authors of
    \cite{DoMa99} show that finitely many Landau-level energies remain infinitely degenerate 
    eigenvalues if $B$ is sufficiently large.
  \item\label{Rem:CoHi}
    Exploiting the existence of spectral gaps of $ H(A,0) $, 
    a Wegner estimate for Landau Hamiltonians with alloy-type random 
    potentials is derived in \cite{CoHi96,BaCoHi97a,BaCoHi97b} which proves that $\nu$ is
    absolutely continuous when restricted 
    to intervals \emph{between} the Landau-level energies. 
    For this result to hold the authors were able to weaken the assumption~(\ref{As:Traeger})  
    on the size of the support
    of the single-site
    potential which our Corollary~\ref{Cor:AT} requires. 
    On the other hand, absolute continuity of $\nu$ at 
    \emph{all} energies is proven in \cite{CoHi96} only 
    for bounded random potentials under the present  
    assumptions on the support. 
  \item\label{Rem:Wa}
    In \cite{Wan97} a Wegner estimate for alloy-type random potentials 
    is derived without assuming a definite sign of the 
    single-site potential. However, this estimate holds only between the 
    Landau-level energies
    for sufficiently strong magnetic field and does not enable 
    one to deduce the (local) existence of the density of states, because it has the ``wrong''
    volume dependence.
  \item 
    In \cite{HuLeWa00} the integrated density of states 
    associated with the \emph{restricted} random Landau Hamiltonian $ P_\ll H(A, V) P_\ll $ 
    of a single but arbitrary Landau level is shown to be absolutely continuous 
    for Gaussian random potentials satisfying the assumptions of Corollary~\ref{Cor:Gauss} (for $ d = 2 $).
  \end{nummer}
\end{remark}
%
%
%
%
%
%
%
\appendix
\section{On Finite-Volume Schr{\"o}dinger Operators with Magnetic Fields}
\label{AppNDI}
%
%
For convenience of the reader (and the authors), this appendix defines 
non-random magnetic Schr{\"o}dinger
operators with Neumann boundary conditions and compiles some of their basic 
properties. In passing, the more familiar basic properties
of the corresponding operators with Dirichlet boundary 
conditions are briefly recalled, see for example \cite{LiMa97,BrHuLe00}.
In particular, we prove a diamagnetic inequality for Neumann Schr{\"o}dinger 
operators and Dirichlet-Neumann bracketing for a wide class
of vector potentials 
including singular ones.  
Altogether, this appendix may be understood to extend some of the results
in the key papers \cite{Kat72,Kat78,AHS78,Sim79MM} to the case of 
Neumann boundary conditions.\\

Throughout this appendix, $ \Lambda \subseteq  \mathbbm{R}^{d}$ denotes a non-empty
 open, not
necessarily proper subset of $ d $-dimensional 
Euclidean space $ \mathbbm{R}^{d} $ with $ d \in \mathbbm{N} $.
Moreover, $ a : \mathbbm{R}^d \to \mathbbm{R}^d$ stands for a vector
potential and $ v :  \mathbbm{R}^d \to \mathbbm{R} $ for a scalar potential
with $v_\pm := \left( |v| \pm v \right)/2 $  denoting its positive
respectively negative part. We will assume throughout that
\begin{equation}\label{eq:ass:av+}
  \left| a \right|^2, \, v_+  
  \in {\rm L}^{1}_{{\rm loc}}( \mathbbm{R}^d ).
\end{equation}
The negative part $v_-$ is assumed to be
a \emph{form perturbation} either of $ H_{\Lambda, \rm N}(a, 0)$ or 
even of $ H_{\Lambda, \rm N}(0, 0)$.
By this we mean that $ v_- $ is 
form-bounded  \cite[Def.\ p.\ 168]{ReSi75} with form bound strictly 
smaller than one 
either relative to $ H_{\Lambda, \rm N}(a, 0)$ or 
even to $ H_{\Lambda, \rm N}(0, 0)$. 
Both operators will be defined in Lemma~\ref{Lem:DefH0} below.
The operator $ H_{\Lambda, \rm N}(0,0) $ is 
the usual Neumann Laplacian, up to a factor of $ -1/2 $.

\begin{remark}
By the diamagnetic inequality, see Prop.~\ref{Prop:Diamag} below, 
we will see that $ v_- $ is a form perturbation of 
$ H_{\Lambda , \rm N} (a, 0) $ 
if it is one 
of $ H_{\Lambda, \rm N}(0,0) $.
If $ \Lambda $ is a bounded open cube, 
an easy-to-check sufficient criterion for $ v_- $ to be even 
infinitesimally form-bounded \cite[Def.\ p.\ 168]{ReSi75}
relative to
$  H_{\Lambda, \rm N}(0,0) $ 
can be taken from \cite[Lemma 2.1]{KirMar82} and reads
\begin{equation}\label{eq:ass:v-}
  v_- \in {\rm L}_{\rm loc}^p(\mathbbm{R}^d)
\end{equation}
with $p =1 $ if $ d = 1 $, some $ p > 1 $ if $d = 2$ and some $ p \geq d/2 $ if $ d \geq 3 $. 
\end{remark} 
\subsection{Definition of magnetic Neumann Schr{\"o}dinger operators}
In a first step, we consider the case $ v = 0 $ and
$ \left| a \right|^2 \in {\rm L}^{1}_{{\rm loc}}(  \mathbbm{R}^d ) $ or,
equivalently, $ a \in \left({\rm L}^{2}_{{\rm loc}}(  \mathbbm{R}^d ) \right)^d $, that is, 
$ a_j \in {\rm L}^{2}_{{\rm loc}}(  \mathbbm{R}^d ) $ for all $ j \in \{1, \dots , d \} $.
We define the sesquilinear form
\begin{equation}  
  h_{\Lambda, {\rm N}}^{a,0}\left(\varphi ,\psi \right):= \frac{1}{2} \, \sum _{j=1}^{d}
  \left\langle \left( {\rm i} \nabla + a\right)_{j} \varphi,\,\,
    \left( {\rm i} \nabla  + a\right)_{j} \psi \right\rangle
\end{equation}
for all $\varphi$ and $\psi$ in its form domain 
\begin{equation} 
  W_a^{1,2}(\Lambda) :=  \left\{ \phi \in {\rm L}^{2}\left( \Lambda \right) : \;
    \left( {\rm i} \nabla + a\right) \phi \in 
    \left( {\rm L}^{2}\left( \Lambda \right) \right)^{d} \right\},
\end{equation}
which might be called a \emph{magnetic Sobolev space}, see 
\cite[Sect. 7.20]{LiLo97} in case $ \Lambda = \mathbbm{R}^d $.
Here and in the following, $ \nabla - {\i} a  $ denotes the gauge-covariant gradient  
in the sense of distributions on $ \mathcal{C}_0^{\infty }\left( \Lambda \right) $.
In particular, this means
\begin{eqnarray}\label{Eq:Dist}
         W_a^{1,2}(\Lambda) =  \bigcap_{j = 1}^d \Big\{ \phi \in {\rm L}^{2}(\Lambda) \; : \;
          \mbox{there is} \;\; \phi_j \in {\rm L}^2(\Lambda) \;\; \mbox{such that} 
          \qquad  \qquad  && \qquad \\[-2ex]
          \langle \, \phi \, , \,  {\rm i} \partial_j \eta + a_j \eta \,\rangle =   
          \langle \, \phi_j \, , \, \eta \,\rangle \quad
          \mbox{for all} \; \eta \in  \mathcal{C}_0^{\infty }\left( \Lambda \right) \Big\}. && \nonumber
\end{eqnarray} 
\begin{remark}
      We emphasize that the condition $ \psi \in  W_a^{1,2}(\Lambda) $ allows for the case that neither
      $\nabla \psi $ nor $ a  \psi $ belongs to $ \left( {\rm L}^{2}\left( \Lambda \right)\right)^d  $.
      In general, $ \psi \in  W_a^{1,2}(\Lambda) $ only implies 
      $  \nabla \psi \in \left( {\rm L}_{\rm loc}^{1}\left( \Lambda \right) \right)^d $ and
      $ \left| \, \psi \, \right| \in  W^{1,2}(\Lambda)  := \big\{ \phi \in {\rm L}^{2}\left( \Lambda \right) : \;
        \nabla \phi \in \left( {\rm L}^{2}\left( \Lambda \right) \right)^{d} \big\} $,
      the usual first-order Sobolev space of ${\rm L}^2$-type. The latter statement is a consequence of 
      the diamagnetic inequality, see Remark~\ref{Rem:Kato} below and \cite{Sim79KI}. 
      If even $ \left| a \right|^2 \in {\rm L}^\infty(\mathbbm{R}^d) $, the magnetic Sobolev space
      coincides with the usual one,
      $ W_a^{1,2}(\Lambda) = W^{1,2}(\Lambda) $, up to equivalence of norms.
\end{remark}
Basic facts about $ h_{\Lambda, {\rm N}}^{a,0} $ are summarized in
\begin{lemma}\label{Lem:DefH0}
  The form $ h_{\Lambda, {\rm N}}^{a,0} $ is densely defined on
  $ {\rm L}^{2}\left(\Lambda \right) $, symmetric, positive and closed.
  It therefore uniquely defines a self-adjoint positive operator  $ H_{\Lambda, {\rm N}}(a,0) $
  on $  {\rm L}^{2}\left(\Lambda \right) $
  which, up to a factor of $ - 1/2 $, is called \emph{magnetic Neumann Laplacian}.
\end{lemma}
\begin{proof}
  Since $  \mathcal{C}_0^{\infty }\left( \Lambda \right) \subset W_a^{1,2}(\Lambda) 
  \subset {\rm L}^{2}\left(\Lambda \right) $
  and $ \mathcal{C}_0^{\infty }\left( \Lambda \right) $ is dense in $ {\rm L}^{2}\left(\Lambda \right) $,
  the form $  h_{\Lambda, {\rm N}}^{a,0} $ is densely defined.
  Its symmetry and positivity are obvious from the definition.
  To prove that $  h_{\Lambda, {\rm N}}^{a,0} $ is also closed we have to show 
  that the space $ W_a^{1,2}(\Lambda) $ is complete with respect to the (metric induced by the form-) norm
  \begin{equation}\label{eq:norm}
    \sqrt{  \left\langle \phi , \, \phi \right\rangle
      + 
      h_{\Lambda, {\rm N}}^{a,0}\left(\phi, \phi \right) }.
  \end{equation}
  To this end,  we proceed along the lines of Sects.~7.20 and 7.3 in \cite{LiLo97} and 
  let $ ( \phi_n )_{n \in \mathbbm{N}} $  be a
  sequence in $ W_a^{1,2}(\Lambda) $ which is Cauchy with respect to the norm (\ref{eq:norm}).
  By completeness
  of ${\rm L}^2(\Lambda)$, there exist functions $ \phi $, $ \psi_j \in {\rm L}^2(\Lambda) $, $j \in \{ 1, \dots , d \}$,
  such that 
  $ \phi_n \to  \phi $ and $ ({\rm i} \nabla + a)_j \phi_n \to \psi_j $ strongly in $ {\rm L}^2(\Lambda) $ as $ n \to \infty $.
  Since $ ({\rm i} \nabla + a)_j  \phi_n \to ({\rm i} \nabla + a)_j \phi $ in the sense of distributions on 
  $ \mathcal{C}_0^\infty(\Lambda) $ as  $ n \to \infty $, we have $  ({\rm i} \nabla + a)_j \phi = \psi_j $ and hence
  $  \phi \in  W_a^{1,2}(\Lambda) $.
 The existence and uniqueness of $ H_{\Lambda, {\rm N}}(a,0) $ follow now from the one-to-one
 correspondence between densely defined, symmetric, bounded below, closed forms and 
 self-adjoint, bounded below operators, see
 \cite[Thm.~VIII.15]{ReSi80}.
\qed
\end{proof}
\begin{remark}
  \begin{nummer}
    \item 
      We recall that the operator $ H_{\Lambda, \rm N}(a,0) $ has the subspace 
      \begin{eqnarray}\label{eq:domain1}
        \mathcal{D}\left(H_{\Lambda, \rm N}(a,0) \right)  :=  \Big\{ \psi \in W_a^{1,2}(\Lambda) \; : \;
          \mbox{there is} \;\; \widetilde\psi \in {\rm L}^2(\Lambda) \;\; \mbox{such that}  \qquad && \\ 
           h_{\Lambda, {\rm N}}^{a,0}\left(\varphi, \psi \right) = \langle \varphi \, , \, \widetilde\psi \,\rangle \quad
          \mbox{for all} \; \varphi \in W_a^{1,2}(\Lambda)  \Big\} && \nonumber
      \end{eqnarray} 
      of its underlying form domain as its operator domain and acts according to 
      $ H_{\Lambda, \rm N}(a,0) \, \psi =  \widetilde\psi $.
    \item \label{Rem:altchar} 
      Let $ D_j\!\left(a\right) $ denote the closure of the symmetric operator 
      $ \mathcal{C}_0^{\infty }\left( \Lambda \right) \ni \psi \mapsto 
      \left({\rm i} \nabla + a\right)_j \psi \in {\rm L}^2(\Lambda)$. Being the closure of a 
      symmetric operator, $ D_j\!\left(a\right) $ is symmetric. 
      The domain of its adjoint $ D_j^\dagger\!\left(a\right) $ is given by
      \begin{equation}\label{eq:domain}
        \mathcal{D}\big( D^{\dagger }_{j} ( a ) \big) := 
        \left\{ \psi \in {\rm L}^{2}\left( \Lambda \right) : \;
          \left({\rm i} \nabla + a \right)_j \psi \in {\rm L}^{2}\left( \Lambda \right) \right\},
      \end{equation}
      because the adjoint of $ \mathcal{C}_0^{\infty }\left( \Lambda \right) \ni \psi \mapsto 
      \left({\rm i} \nabla + a\right)_j \psi $ coincides with that of its closure.
      While for a proper subset $\Lambda\neq \mathbbm{R}^d $ the operator $D_j \!\left(a\right)$ 
      is not self-adjoint, 
      it is so for $ \Lambda = \mathbbm{R}^d $ \cite[Lemma~2.5]{Sim79MM}. In the latter case it
      may physically be interpreted, up to a sign,  
      as the $j^{\rm th}$
      component of the velocity (operator).
      By construction the magnetic Neumann Laplacian is a form sum of $d$ operators in accordance with
      \begin{equation}\label{eq:neumann}
        H_{\Lambda, \rm N}(a,0) = \frac{1}{2} \,
        \sum_{j=1}^{d} D_{j}\!\left(a\right) D_{j}^{\dagger }\!\left(a\right),
      \end{equation}
      where the self-adjoint positive operator 
      $D_{j}\!\left( a\right) D_{j}^{\dagger }\!\left( a\right)  $
      comes from the closed form $ \big\langle D_{j}^{\dagger }\!\left(a\right) \varphi ,\,\, 
        D_{j}^{\dagger }\!\left(a\right) \psi \big\rangle  $
      with form domain (\ref{eq:domain}). Note that (\ref{eq:domain}) is just the $j^{\rm th}$
      set of the intersection on the r.h.s. of (\ref{Eq:Dist}).
      See also Thm.\ X.25 in \cite{ReSi75}.
    \item 
      Restricting the form $ h_{\Lambda , {\rm N}}^{a,0} $ to the domain $  \mathcal{C}_0^{\infty }\left( \Lambda \right) 
      \subset W_a^{1,2}(\Lambda) $, 
      one obtains a form which is closable in $ W^{1,2}_a(\Lambda) $ 
      with respect to the norm (\ref{eq:norm}), see \cite{Sim79MM,LiMa97,BrHuLe00}. 
      Its closure $ h_{\Lambda , {\rm D}}^{a,0} $ is uniquely associated
      with another self-adjoint positive operator $ H_{\Lambda, {\rm D}}(a,0) $ on $  {\rm L}^{2}\left(\Lambda \right) $
      which, up to a factor of $ - 1/2 $, is called \emph{magnetic Dirichlet Laplacian}.
      For general $  a  \in \left( {\rm L}_{{\rm loc}}^2( \mathbbm{R}^d ) \right)^d $
      the space $ \mathcal{C}_0^\infty(\Lambda) $ is not contained in $ \mathcal{D}\left(H_{\Lambda, \rm N}(a,0) \right) $,
      see (\ref{eq:domain1}).
      As a consequence, $  H_{\Lambda, {\rm N}}(a,0) $ in general cannot be restricted to $ \mathcal{C}_0^\infty(\Lambda) $.
      This stands in contrast to the case $ a = 0 $ where 
      $ H_{\Lambda, {\rm D}}(0,0) $ is the \emph{Friedrichs extension} of the restriction of $ H_{\Lambda, {\rm N}}(0,0) $
      to $ \mathcal{C}_0^\infty(\Lambda) $.
      As the Dirichlet counterpart of (\ref{eq:neumann}) we only have the inequality  
      $  H_{\Lambda, \rm D}(a,0) \leq \frac{1}{2} 
      \sum_{j=1}^{d} D_{j}^{\dagger }\!\left(a\right) D_{j}\!\left(a\right) $ which is
      meant in the sense of forms \cite[Def. on p.~269]{ReSi78}. 
      The operators $ H_{\mathbbm{R}^d\! , \rm N}(a,0) $ and 
      $ H_{\mathbbm{R}^d\! , \rm D}(a,0) $ are equal,
      see \cite{Sim79MM}.
    \item In the \emph{free} case, which is characterized by $a=0$ and $ v = 0 $, the just defined operators 
      $H_{\Lambda,\rm D}(0, 0)$ and $H_{\Lambda, \rm N}(0, 0)$ coincide, up to a
      factor of $-1/2$, with the usual Dirichlet- and Neumann-Laplacian \cite[p. 263]{ReSi78}, respectively. 
 \end{nummer}
\end{remark}
In a second and final step, we let $v_+ \in {\rm L}_{{\rm loc}}^1( \mathbbm{R}^d ) $ and
assume $ v_- $ to be a form perturbation of  
$ H_{\Lambda , \rm N} (a, 0) $.
As a consequence, the sesquilinear form 
\begin{equation}
  h_{\Lambda, \rm N}^{a, v}\left(\varphi , \psi \right) := 
  h_{\Lambda, \rm N}^{a, 0}\left(\varphi , \psi \right) + 
   \left\langle  v_+^{1/2} \varphi ,  
     v_+^{1/2} \psi \right\rangle -
    \left\langle v_-^{1/2} \varphi ,  
      v_-^{1/2} \psi \right\rangle
\end{equation}
is well defined for all $ \varphi $ and $ \psi $ in its form domain  
$ \mathcal{Q}\big( h_{\Lambda, \rm N}^{a, v} \big)
 := W_a^{1,2}(\Lambda) \cap \mathcal{Q}\left( v_+ \right) $, where
\begin{equation}
 \mathcal{Q}\left(v_+ \right)
 := \left\{ \phi \in {\rm L}^2\left( \Lambda \right) : \,
   v_+^{1/2} \phi \in  {\rm L}^2\left( \Lambda \right) \right\}. 
\end{equation}
Basic facts about $ h_{\Lambda, \rm N}^{a, v} $ are summarized in
\begin{lemma}\label{Lem:Defhav}
  The form  $ h_{\Lambda, \rm N}^{a, v} $ is
  densely defined  on $ {\rm L}^2\left( \Lambda \right) $,
  symmetric, bounded below and closed.
  It therefore uniquely defines a self-adjoint, bounded below 
  operator $ H_{\Lambda, \rm N}(a, v) $ on $ {\rm L}^2\left( \Lambda \right) $ 
  which is called
  \emph{magnetic Neumann Schr{\"o}dinger operator}.
\end{lemma} 
\begin{proof}
  The  domain 
  $ W_a^{1,2}(\Lambda) \cap \mathcal{Q}\left( v_+ \right) $ of $  h_{\Lambda, \rm N}^{a, v_+} $
  is dense in 
  $ {\rm L}^2\left( \Lambda \right) $, because both $ W_a^{1,2}(\Lambda) $ and  $ \mathcal{Q}\left( v_+ \right) $
  contain $ \mathcal{C}_0^\infty(\Lambda) $.
  Hence $ H_{\Lambda , \rm N} (a, v_+) $ is well defined as a form sum of $  H_{\Lambda , \rm N} (a, 0) $
  and $ v_+ $. 
  Moreover, $  h_{\Lambda, \rm N}^{a, v_+} $ is symmetric, positive and closed,
  because it is the sum of two of such forms.
  Since $ H_{\Lambda , \rm N} (a, 0) \leq H_{\Lambda , \rm N} (a, v_+) $, the negative part
  $ v_- $ of $v$ is also 
  a form perturbation of $ H_{\Lambda , \rm N} (a, v_+) $.
  The proof of the lemma is then completed by the KLMN-theorem \cite[Thm.\ X.17]{ReSi75}.
\qed
\end{proof}
\begin{remark}
      Since the form domain of $h_{\Lambda, \rm D}^{a,0} $ is contained in $ W_a^{1,2}(\Lambda) $,
      the negative part $ v_- $ of $ v $
      is also a form perturbation of $ H_{\Lambda, \rm D}(a, 0) \leq H_{\Lambda, \rm D}(a, v_+) $. 
      Hence one may apply the KLMN-theorem to define, similarly 
      to $ H_{\Lambda, \rm N}(a,v)$, what is called the \emph{magnetic Dirichlet Schr{\"o}dinger operator}
      and denoted as $ H_{\Lambda, \rm D}(a,v) $.
\end{remark}
An immediate consequence of the definition of $ H_{\Lambda, \rm X}(a, v) $ is the fact that so-called
\emph{decoupling} and \emph{Dirichlet-Neumann bracketing} continues to hold for $ a \neq 0 $
as in the case $ a = 0 $, see Props.~3 and 4 in Sect.\ XIII.15 of \cite{ReSi78}, and 
\cite{CoScSe78,Nak00} 
for smooth $ a \neq 0 $.
\begin{proposition}
  Let ~$\left| a \right|^2$, $v_+ \in {\rm L}_{\rm
    loc}^{1} ( \mathbbm{R}^d ) $ and ~$v_-$ be a form perturbation of 
  $ H_{\Lambda, \rm N }(a,0) $. 
  Moreover, let $ \Lambda_1 , \Lambda_2 \subset \mathbbm{R}^d $
  be a disjoint pair of non-empty open sets.
  \begin{indentnummer*}
    \item
      Then the orthogonal decomposition
      \begin{equation}\label{eq:MagnDMdecoupling}
        H_{\Lambda_1\cup\Lambda_2, \rm X}\left( a, v\right) =
        H_{\Lambda_1, \rm X}\left( a, v\right) \oplus
        H_{\Lambda_2, \rm X}\left( a, v\right) 
      \end{equation}
      holds for both $\rm X = \rm D$ and $\rm X= \rm N$ on $ {\rm L}^2\left(\Lambda_1 \cup \Lambda_2 \right)=
      {\rm L}^{2}\left( \Lambda_1 \right)\oplus {\rm L}^{2}\left( \Lambda_2 \right)$. 
    \item
      Let $\Lambda :=\overline{\Lambda_1\cup\Lambda_2}^{\, \rm{int}}$ be defined as
      the interior of the closure 
      of the union of $ \Lambda_1 $ and $ \Lambda_2 $, and suppose that the interface 
      $ \Lambda\setminus(\Lambda_1\cup\Lambda_2) $ is of
      $d$-dimensional Lebesgue measure zero. Then the inequalities  
      \begin{equation}\label{eq:MagnDMbracketing}
        H_{\Lambda_1\cup\Lambda_2, \rm N}\left( a, v\right) \leq
        H_{\Lambda , \rm N}\left( a, v\right) \leq
        H_{\Lambda , \rm D }\left( a, v\right) \leq
        H_{\Lambda_1\cup\Lambda_2, \rm D}\left( a, v\right)
      \end{equation}
      hold in the sense of forms.
    \end{indentnummer*}
\end{proposition}
\begin{proof}
        The proofs of Props.~3 and 4 in Sect.\ XIII.15 of \cite{ReSi78}
        for the free case carry over to the case $ a \neq 0 $ and $ v \neq 0 $. In particular, 
        the inclusion relations 
        between the various form domains for $ a = 0 $ and $ v = 0 $ hold analogously for the 
        form domains in the case $ a \neq 0 $ and $ v \neq 0 $.
        \qed
\end{proof}
%
%
\subsection{Diamagnetic inequality}
A useful tool in the study of Schr{\"o}dinger operators with magnetic fields is  
\begin{proposition}\label{Prop:Diamag}
  Let ~$ \Lambda \subseteq \mathbbm{R}^d $ be open, $\left| a \right|^2$, $v_+ \in {\rm L}_{\rm
    loc}^{1} ( \mathbbm{R}^d ) $ and ~$v_-$ be a form perturbation of 
  $ H_{\Lambda, \rm N }(0,0) $. Then ~$v_-$ is a form perturbation of 
  $ H_{\Lambda, \rm N }(a,0) $ with form bound not exceeding the one for $ a = 0 $
  and the inequality
  \begin{equation}\label{eq:NDI}
    \big|\, {\rm e}^{-t\, H_{\Lambda , \rm X}\left(a,v\right) }\psi \big| \leq
    {\rm e}^{-t\, H_{\Lambda , \rm X} \left(0 ,v\right) } \left| \psi \right|
   \end{equation}
   holds for all $ \psi\in{\rm L}^2\left(\Lambda\right) $, all $ t\geq 0 $ and both 
   $ \rm X = \rm D $ and $ \rm X = \rm N $ .
\end{proposition}
\begin{remark}
  \begin{nummer}
    \item
      For the  Dirichlet version $ \rm X = \rm D $ of the \emph{diamagnetic inequality} (\ref{eq:NDI})
      to hold, 
      it would be sufficient that $ v_- $ is a form perturbation of
      $ H_{\Lambda, \rm D}(0,0) $.
    \item 
      Inequality (\ref{eq:NDI}) for $ \Lambda = \mathbbm{R}^d $
      dates back to \cite{Kat72,Sim77,HeScUh77,Kat78,AHS78,Sim79KI,Sim79MM}.
      It is also known to hold for $ \Lambda \neq \mathbbm{R}^d $ and $ \rm X = \rm D $, 
      even under the weaker assumptions
      $\left| a \right|^2$, $v_+ \in {\rm L}_{\rm loc}^{1}\left( \Lambda \right) $, 
      see \cite{PeSe81,LiMa97}. These assumptions still guarantee 
      that the operators $ H_{\Lambda, \rm D}(a,v) $ and $ H_{\Lambda, \rm N}(a,v) $ 
      are definable as self-adjoint 
      operators via forms. 
      However, for arbitrary open $ \Lambda \neq \mathbbm{R}^d $
      the proof of (\ref{eq:NDI}) for $ \rm X = \rm N $ would be more complicated 
      than the one which we will give under the stronger assumptions of Prop.~\ref{Prop:Diamag}.
      The reason is that a gauge function more fancy than that in Lemma~\ref{Lemma:wegeich}
      would be needed in order to avoid integration of $ a_j $ across the boundary of $ \Lambda $.
      For a ``simply shaped'' $ \Lambda $, like a cube, such complications do not arise
      which implies that our proof would go through for cubes under the weaker assumptions.
    \item
      If ~$ a=0 $ inequality (\ref{eq:NDI}) is equivalent to the assertion that 
      $ H_{\Lambda , \rm X} \left(0 ,v\right) $ is the (negative of the) generator of a
      positivity-preserving one-parameter operator semigroup on $ {\rm L}^2(\Lambda) $, 
      see \cite[pp.~186]{ReSi75}.
      For general $ a \in \left( {\rm L}_{\rm loc}^2(\mathbbm{R}^d) \right)^d $
      inequality (\ref{eq:NDI}) asserts that the semigroup generated by $  H_{\Lambda, \rm X}(0,v) $
      dominates the one generated by $  H_{\Lambda, \rm X}(a,v) $.
    \item\label{Rem:Kato}
      It follows from \cite{HeScUh77,Sim79KI} that 
      (\ref{eq:NDI}) is equivalent
      to the following pair of statements: 
      \begin{eqnarray}
        \mbox{\rm (a)} && \;\; \psi \in \mathcal{D}\big(  H_{\Lambda, \rm X}(a,v) \big) \;\;\; \mbox{implies} \;\;\;  
         \left| \psi \right| \in \mathcal{Q}\big( h_{\Lambda , \rm X}^{0,v} \big), 
         \nonumber \\
        \mbox{\rm (b)} && \;\; h_{\Lambda, \rm X}^{0,v}( \varphi ,\left| \psi \right|)
         \leq \Re \left\langle \varphi \, \sgn \psi \, , \,  
           H_{\Lambda, \rm X}(a,v) \, \psi \right\rangle 
         \nonumber \\
         && \mbox{for all $ \varphi \in \mathcal{Q}\big( h_{\Lambda , \rm X}^{0,v} \big) $ with
          $ \varphi \geq 0 $ and all
          $ \psi \in \mathcal{D} \big(  H_{\Lambda, \rm X}(a,v) \big) $, } \qquad \qquad \nonumber 
      \end{eqnarray}
      where the \emph{signum function} associated with $ \psi $ is defined by 
      $ \left(\sgn {\psi}\right)(x) :=  \psi(x) / | \psi(x) | \in \mathbbm{C}$ 
      if $ \psi (x) \neq 0 $ and zero otherwise. 
      If $ a = 0 $ these statements boil down to a  
      Beurling-Deny criterion \cite[Thm.~1.3.2]{Dav90}
      for $  H_{\Lambda, \rm X}(0,v) $ which guarantees 
      that it generates a positivity-preserving semigroup.
      Inequality~(b) with $ \rm X = \rm N $ and $ v = 0 $ basically corresponds 
      to the germinal \emph{distributional inequality of Kato}, which
      he proved~\cite{Kat72} for $ a \in \left(\mathcal{C}^1(\mathbbm{R}^d)\right)^d $. 
      In case $ \Lambda \neq \mathbbm{R}^d $ and $ \rm X = \rm N $, 
      we are not aware of a reference proving (\ref{eq:NDI}) 
      or (a) and (b) for singular \nolinebreak $a$.
    \end{nummer}
\end{remark}
Our proof of the diamagnetic inequality (\ref{eq:NDI}) for $ \rm X = \rm N $ 
will mimic the proof in~\cite{Sim79MM}, 
where the case $ \Lambda = \mathbbm{R}^d $ is considered, 
see also Sect.~1.3 in \cite{CyFr87}.  It relies on the fact that for one dimension the
vector potential can be removed by a gauge transformation.
More precisely, for each $ j \in \{ 1 , \dots , d \} $ the operator $ D^{\dagger }_{j}\!\left(a\right) $ 
is unitarily equivalent to $ D^{\dagger }_{j}\!\left( 0\right) $.
\begin{lemma}\label{Lemma:wegeich}
  Let ~$\left| a \right|^2 \in {\rm L}_{\rm loc}^{1} ( \mathbbm{R}^d ) $ 
  and define a \emph{(gauge)} function $ \lambda_j: \mathbbm{R}^d \to \mathbbm{R} $ through
  \begin{equation}
    \lambda_j\left( x \right) := \int_{0}^{x_{j}}\! \d y_j\,\,
    a_{j}\left( x_{1},\ldots , x_{j-1},y_j,x_{j+1},\ldots , x_{d} \right).
  \end{equation}
  For open $\Lambda \subseteq \mathbbm{R}^d $
  it induces a densely defined and self-adjoint 
  multiplication operator $ \lambda_j $ on ${\rm L}^2\left( \Lambda  \right) $.
  The corresponding unitary operator ${\rm e}^{ - {\rm i}\lambda_j} $ maps
  $  \mathcal{D}\big( D^{\dagger }_{j} ( a ) \big) $ onto 
  $  \mathcal{D}\big( D^{\dagger }_{j} ( 0 ) \big) $, recall {\rm (\ref{eq:domain})}, and one has
  \begin{equation}\label{eq:oneDgaugeInvariance}
    D_j^\dagger\!\left(a\right) \psi = 
    {\rm e}^{ {\rm i}\lambda_j}  D_j^\dagger\!\left(0\right)
    {\rm e}^{ - {\rm i}\lambda_j} \psi
  \end{equation}
  for all $ \psi \in  \mathcal{D}\big( D^{\dagger }_{j} ( a ) \big) $.
\end{lemma}
\begin{proof}
  Fubini's theorem and the Cauchy-Schwarz inequality show that $ \lambda_j \in
  {\rm L}_{{\rm loc}}^2(\mathbbm{R}^d ) $. 
  Therefore, the induced multiplication operator on its maximal domain
  $ \mathcal{D} ( \lambda_j ) := \big\{ \psi \in {\rm L}^2\left( \Lambda  \right) : 
    \,  \lambda_j  \psi \in {\rm L}^2\left( \Lambda  \right) \big\} \supset 
  \mathcal{C}_0^\infty\left( \Lambda \right) $ is densely defined and self-adjoint.
  Moreover, 
  since $\psi \in \mathcal{D}\big( D^{\dagger }_{j} ( a ) \big) $
  implies $ \nabla_j \psi \in {\rm L}_{\rm loc}^1\left(\Lambda\right) $, we are allowed
  to use the product and chain rule for distributional derivatives \cite[pp. 150]{GT}
  which yield 
  $
    \nabla_j \left( {\rm e}^{ - {\rm i}\lambda_j} \psi \right) 
  $\hspace{0pt}$=
    {\rm e}^{ - {\rm i}\lambda_j}  \nabla_j  \psi -  
    {\rm e}^{ - {\rm i}\lambda_j} {\rm i} a_j \psi
  $.
\qed
\end{proof}
\begin{proof}[of Prop.~\ref{Prop:Diamag}]
  For $ \rm X = \rm D $ see \cite{PeSe81,LiMa97,BrHuLe00}. 
  The proof for $ \rm X = \rm N $ consists of three steps. 

  In the first step, we assume
  $ v \in {\rm L}_{\rm loc}^1 ( \mathbbm{R}^d ) $ to
  be bounded from below.
  In this case $H_{\Lambda, \rm N}(a, v)$ is a form sum of $d+1$
  operators each of which is bounded from below, 
  recall Remark~\ref{Rem:altchar} and Lemma~\ref{Lem:Defhav}.
  Hence we may employ the
  strong Lie-Trotter product formula generalized to form sums of several operators
  \cite{KatMas78} and write
  \begin{equation}\label{eq:KMTpf}
    \e^{-t H_{\Lambda, \rm N} \left(a, v\right)} =
    \mathop{\mbox{s-lim}}_{n\to \infty }\left(
      \e^{-t D_1\left(a\right) D_1^\dagger\!\left(a\right) /2n} \cdots 
      \e^{-t D_d\left(a\right) D_d^\dagger\left(a\right) /2n} \e^{-t v/n}
    \right)^{n}.
  \end{equation}
  Gauge equivalence (\ref{eq:oneDgaugeInvariance}) now shows that
  \begin{equation}\label{eq:oneDgaugeInv}
    \e^{-t D_j\!\left(a\right) D_j^\dagger\!\left(a\right) /2n} =
    \e^{{\i} \lambda_j} 
    \e^{-t D_j\!\left(0\right) D_j^\dagger\!\left(0\right) /2n}
    \e^{- {\i}\lambda_j}
  \end{equation}
  for all $j \in \left\{ 1, \dots , d \right\}$ and all $t\geq0$.
  By the distributional inequality 
  $\big| \nabla_j  \left| \psi \right| \big| \leq \left| \nabla_j \psi \right| $,
  valid for all $ \psi \in  \mathcal{D}\big( D^{\dagger }_{j} ( 0 ) \big) $ 
  \cite[Thm.~6.17]{LiLo97}, the operator 
  $  D_j\!\left(0\right)D_j^\dagger\!\left(0\right) $ obeys a
  Beurling-Deny criterion \cite[Thm.~1.3.2]{Dav90}
  and hence is the 
  generator 
  of a positivity-preserving semigroup.
  It follows that
  \begin{equation}\label{eq:oneDsemiGrDom}
    \left| 
      {\rm e}^{-tD_j\left(a\right) D_j^\dagger\!\left(a\right) /2n} \psi  
    \right| \leq 
    {\rm e}^{-tD_j\!\left(0\right) D_j^\dagger\!\left(0\right) /2n}
    \left| \psi \right| 
  \end{equation}
  for all $\psi \in {\rm L}^2\left(\Lambda\right)$ and
  all $t\geq 0 $.
  This together with (\ref{eq:KMTpf}) implies the assertion (\ref{eq:NDI}) (with $ \rm X = \rm N $)
  for scalar potentials $ v \in {\rm L}_{\rm loc}^1( \mathbbm{R}^d ) $
  which are bounded from below.

  In the second step,
  we prove that if $ v_- $ is a form perturbation of $ H_{\Lambda, \rm X}(0,0) $
  then it is also one of $ H_{\Lambda , \rm X} (a, 0) $ with form bound not exceeding the one 
  for $ a = 0 $ (see \cite{AHS78} or \cite[Thm.~15.10]{Sim79} 
  for the case $ \Lambda = \mathbbm{R}^d $).
  This follows from (\ref{eq:2DMIresPow}) below with $ v = 0 $ and $ \alpha = 1/2 $
  together with 
  the fact that the form bound of $ v_- $ relative to
  $  H_{\Lambda , \rm X} (a, 0) $ can be expressed as
  \begin{equation}
    \lim_{E\to \infty } \left\| \left( H_{\Lambda, \rm X}(a,0) + E \right)^{-1/2} v_- 
      \left( H_{\Lambda, \rm X}(a,0) + E \right)^{-1/2} \right\|,
  \end{equation}
  see \cite[Prop.~1.3(ii)]{CyFr87}. 
  Here $ \left\| \cdot \right\| $ denotes the (uniform) norm of bounded operators on $ {\rm L}^2(\Lambda) $.

  In the third step,
  we extend the validity of (\ref{eq:NDI}) (with $ \rm X = \rm N $) to 
  scalar potentials $v$ with $ v_+ \in {\rm L}_{\rm loc}^1(\mathbbm{R}^d) $ and 
  $v_-$ being a form perturbation of $ H_{\Lambda, \rm N}(0, 0) $. 
  To this end, we approximate $v$ by $v_n$ defined through 
  $v_n\left(x\right):=\max\left\{-n, v\left(x\right)\right\}$, $ x \in \mathbbm{R}^d$ , $n\in\mathbbm{N}$.  
  Monotone convergence for forms \cite[Thm.~S.16]{ReSi80} yields the convergence of
  $H_{\Lambda , \rm N}\left(a,v_n\right)$ to
  $H_{\Lambda , \rm N}\left(a,v\right)$ in the strong resolvent sense as
  $n\to\infty$. It follows that
  \begin{equation}
    \mathop{\mbox{s-lim}}_{n\to\infty} \, \e^{-t H_{\Lambda , \rm N}\left(a,v_n\right)} =  
    \e^{ -t H_{\Lambda , \rm N}\left(a,v\right)}
  \end{equation}
  for all $t \geq 0$. Since
  (\ref{eq:NDI}) (with $ \rm X = \rm N $)
  holds for each $v_n$ by the first step, the proof is complete. 
  \qed
\end{proof}

\subsection{Some consequences}\label{sec:consequences}
We list some immediate consequences of the diamagnetic inequality.
For this purpose, we assume the situation of Prop.~\ref{Prop:Diamag}.
\begin{nummer}
  \item
    Powers of the resolvent of the self-adjoint operator $ H_{\Lambda , \rm X}(a,v) $
    may be expressed in terms of its semigroup by using the
    functional calculus. This gives the
    integral representation 
    \begin{equation}\label{eq:ressemi}
      \left(H_{\Lambda , \rm X}\left(a,v\right) -z\right)^{-\alpha}=
      \frac{1}{\left( \alpha -1 \right)!} \,
      \int_0^\infty \!\d t\, t^{\alpha-1} \, {\rm e}^{tz} \,  
      {\rm e}^{-t H_{\Lambda , \rm X}\left(a,v\right)},
    \end{equation}
    which is valid for  all $\alpha>0$, all $ z \in \mathbbm{C} $ with 
    $\Re z < \infspec H_{\Lambda , \rm X}\left(a,v\right)$ and 
    both $ \rm X = \rm D $ and $ \rm X = \rm N $.
    Here $ \alpha \mapsto \left( \alpha -1 \right)! $ denotes Euler's gamma function~\cite{GrRy}.
    Inequality (\ref{eq:NDI})
    then implies the {\em diamagnetic inequality for powers of the resolvent}
    \begin{equation}\label{eq:2DMIresPow}
      \big| \, \left(H_{\Lambda , \rm X}\left(a,v\right) - z \right)^{-\alpha} \psi \big| \leq 
      \left(H_{\Lambda , \rm X}\left(0,v\right) - \Re z \right)^{-\alpha} \left| \psi \right|,
    \end{equation}
    valid for all $\psi\in{\rm L}^2\left(\Lambda\right)$ and all $z \in \mathbbm{C}$ 
    with $\Re z < \infspec H_{\Lambda , \rm X}\left(0,v\right) $. 
    We recall \cite{Sim76} that the ground-state energy goes up when the magnetic field is turned on, in symbols,
    $ \infspec H_{\Lambda , \rm X}\left(0,v\right) \leq \infspec H_{\Lambda , \rm X}\left(a,v\right)$.
    This follows from Remark~\ref{Rem:Kato}(b) or inequality (\ref{eq:diazusum}) below if its r.h.s. is finite.
  \item
    If $ H_{\Lambda, \rm X}(0,v) $ has purely discrete spectrum or, 
    equivalently \cite[Thm. XIII.64]{ReSi78}, has compact resolvent,
    the Dodds-Fremlin-Pitt theorem \cite[Thm.~2.2]{AHS78} together with (\ref{eq:2DMIresPow})
    implies that $ H_{\Lambda, \rm X}(a,v) $ has also compact resolvent and 
    hence purely discrete spectrum. 
    In turn, $ H_{\Lambda , \rm X}(0, v) $ has purely discrete spectrum if the free operator
    $ H_{\Lambda, \rm X}(0,0) $ has and if $ v $ is a form perturbation of
    $ H_{\Lambda, \rm X}(0,0) $ \cite[Thm.~XIII.68]{ReSi78}.
    While $ H_{\Lambda, \rm D}(0,0) $ has purely discrete spectrum for arbitrary bounded open 
    $ \Lambda \subset \mathbbm{R}^d $, $ H_{\Lambda, \rm N}(0,0) $ only has if $ \Lambda $
    possesses an additional property, for example the \emph{segment property}, see \cite[pp.~255]{ReSi78}.
    For example, if $ \Lambda $ is a bounded open \emph{cube} the spectra of 
    $ H_{\Lambda, \rm D}(a,-v_-) $ and $ H_{\Lambda, \rm N}(a,-v_-) $ are both purely discrete.
    Moreover, by the min-max principle the addition of the positive multiplication operator $ v_+ $ to
    $ H_{\Lambda, \rm X}(a, -v_-) $ cannot 
    create essential spectrum. As a consequence, $  H_{\Lambda, \rm X}(a, v) $ has purely discrete spectrum
    for both $ \rm X = \rm D $ and $ \rm X = \rm N $ if $ \Lambda $ is a bounded open cube.
  \item
    The diamagnetic inequality (\ref{eq:NDI})
    together with Lemma~15.11 in \cite{Sim79} implies the
    \emph{diamagnetic inequality for partition functions}
    \begin{equation}\label{eq:diazusum}
      \Tr\left[ \e^{ -t H_{\Lambda , \rm X}(a, v)}  \right]
        \leq
        \Tr\left[ \e^{ -t H_{\Lambda , \rm X}(0, v)}  \right]
    \end{equation}
    for all $ t > 0 $ and both $\rm X = \rm D $ and $\rm X = \rm N $, provided that the r.h.s. is finite.
    The latter is the case if $ \Lambda $ is a bounded open cube, for example.
    This follows from Dirichlet-Neumann bracketing (see (\ref{eq:MagnDMbracketing}) with $ a = 0 $),
    the facts that  $ v_+ \geq 0 $ and $ v_- $ is a form perturbation of $  H_{\Lambda , \rm N}(0, 0) $,
    and the finiteness of the free Neumann partition function (see \cite[Prop.~2.1(c)]{KirMar82} or (\ref{eq:zudoof})).
\end{nummer}

%
\begin{acknowledgement}
\addcontentsline{toc}{section}{Acknowledgement}
It is a pleasure to thank Kurt Broderix, Dirk Hundertmark, 
Thomas Hoffmann-Ostenhof and Georgi D. Raikov
for helpful remarks and stimulating discussions.
This work was supported by the Deutsche Forschungsgemeinschaft under
grant nos. Le 330/10  and Le 330/12. The latter is a project within the
Schwerpunktprogramm ``Interagierende stochastische Systeme von hoher
Komplexit{\"a}t'' (DFG Priority Programme SPP 1033).
\end{acknowledgement}%

\begin{noteadd}
  After submission of the present paper we learned of the interesting 
  paper \emph{The $L^p$-theory of the spectral shift function,
  the Wegner estimate, and the integrated density
  of states for some random operators},
  Commun. Math. Phys. {\bf{218}}, 113--130 (2001), by J.\ M.\ Combes, P.\ D.\ Hislop
  and S.\ Nakamura.
  Among other things, their approach yields Wegner estimates for rather
  general magnetic fields and certain bounded random potentials. While 
  these estimates do not imply absolute continuity of the integrated 
  density of states, they yield H\"older continuity of arbitrary 
  order strictly smaller than one.
  The recent preprint \emph{The integrated density of states for some random operators with nonsign 
    definite potentials}, mp\_arc 01-139 (2001), by P.\ D.\ Hislop and F.\ Klopp extends part of this 
  result to single-site potentials taking values of both signs. 
\end{noteadd}
%


%

%
%

\begin{thebibliography}{MM}
%
\addcontentsline{toc}{section}{References}
\frenchspacing
%

\bibitem{Adl81} 
  \au{R.J.}{Adler}:
  \bti{The geometry of random fields}
  \pub{Wiley}{Chichester}{1981}

\bibitem{AnFoSt82}
  \au{T.}{Ando}, \au{A.B.}{Fowler}, \au{F.}{Stern}:
  \ti{Electronic properties of two-dimensional systems}
  \z{Rev. Mod. Phys.}{54}{437--672}{1982}

\bibitem{AHS78} 
  \au{J.}{Avron}, \au{I.}{Herbst}, \au{B.}{Simon}:
  \ti{Schr{\"o}dinger operators with magnetic fields. I. General interactions}
  \z{Duke Math. J.}{45}{847--883}{1978}


\bibitem{BaCoHi97a} 
  \au{J.-M.}{Barbaroux}, \au{J.M.}{Combes}, \au{P.D.}{Hislop}:
  \ti{Localization near band edges for random Schr{\"o}dinger operators}
  \z{Helv. Phys. Acta}{70}{16--43}{1997}

\bibitem{BaCoHi97b} 
  \au{J.-M.}{Barbaroux}, \au{J.M.}{Combes}, \au{P.D.}{Hislop}:
  \ti{Landau Hamiltonians with unbounded random potentials}
  \z{Lett. Math. Phys.}{40}{335--369}{1997}


\bibitem{Bau92}
   \au{H.}{Bauer}:
   \bti{Ma{\ss}- und Integrationstheorie}
   $2$. Auflage,
   \pub*[in German]{de Gruyter}{Berlin}{1992}
   English translation to appear





\bibitem{BoEn84}
  \au{V.L.}{Bonch-Bruevich}, \au{R.}{Enderlein}, \au{B.}{Esser},
  \au{R.}{Keiper}, \au{A.G.}{Mironov}, \au{I.P.}{Zvyagin}:
  \bti{Elektronentheorie ungeordneter Halbleiter}
  \pub*[in German. Russian original: \pub{Nauka}{Moscow}{1981}]%
  {VEB Deutscher Verlag der Wissenschaften}{Berlin}{1984}





\bibitem{BrHuLe93} 
  \au{K.}{Broderix}, \au{D.}{Hundertmark}, \au{H.}{Leschke}:
  \ti{Self-averaging, decomposition and asymptotic properties of the
    density of states for random Schr{\"o}dinger operators with constant
    magnetic field} 
  In:
  \bti{Path integrals from meV to MeV: Tutzing '92}
  Grabert, H., Inomata, A., Schulman, L.S., Weiss, U.  (eds.),
  \pub[pp. 98--107]{World Scientific}{Singapore}{1993}


\bibitem{BrHuLe00} 
  \au{K.}{Broderix}, \au{D.}{Hundertmark}, \au{H.}{Leschke}:
  \ti{Continuity properties of Schr{\"o}dinger semigroups with magnetic
    fields} 
  \z{Rev. Math. Phys.}{12}{181--225}{2000}

\bibitem{CaLa90}
  \au{R.}{Carmona}, \au{J.}{Lacroix}:
  \bti{Spectral theory of random Schr{\"o}dinger operators}
  \pub{Birkh{\"a}user}{Boston}{1990}


\bibitem{CoHi94} 
  \au{J.M.}{Combes}, \au{P.D.}{Hislop}:
  \ti{Localization for some continuous, random Hamiltonians in 
    $d$-dimensions}
  \z{J. Funct. Anal.}{124}{149--180}{1994}


\bibitem{CoHi96} 
  \au{J.M.}{Combes}, \au{P.D.}{Hislop}:
  \ti{Landau Hamiltonians with random potentials: Localization and the
    density of states}
  \z{Commun. Math. Phys.}{177}{603--629}{1996}

\bibitem{CoHiMo96} 
  \au{J.M.}{Combes}, \au{P.D.}{Hislop}, \au{E.}{Mourre}:
  \ti{Spectral averaging, perturbation of singular spectra, and localization}
  \z{Trans. Am. Math. Soc.}{348}{4883--4894}{1996}

\bibitem{CoScSe78}
   \au{J.M.}{Combes}, \au{R.}{Schrader}, \au{R.}{Seiler}:
   \ti{Classical bounds and limits for energy distributions of Hamilton operators in
        electromagnetic fields}
   \z{Ann. Phys. (N.Y.)}{111}{1--18}{1978}

\bibitem{CrSi83}
  \au{W.}{Craig}, \au{B.}{Simon}:
  \ti{Log H{\"o}lder continuity of the integrated density of states for stochastic Jacobi
    matrices}
  \z{Commun. Math. Phys.}{90}{207--218}{1983}


\bibitem{CyFr87}
  \au{H.L.}{Cycon}, \au{R.G.}{Froese}, \au{W.}{Kirsch}, \au{B.}{Simon}: 
  \bti{Schr{\"o}dinger operators}
  \pub{Springer}{Berlin}{1987}


\bibitem{Dav90}
  \au{E.B.}{Davies}:
  \bti{Heat kernels and spectral theory}
  Paperback edition,
  \pub{Cambridge Univ. Press}{Cambridge}{1990}


\bibitem{DelSou84}
  \au{F.}{Delyon}, \au{B.}{Souillard}:
  \ti{Remark on the continuity of the density of states of 
    ergodic finite difference operators}
  \z{Commun. Math. Phys.}{94}{289--291}{1984}






\bibitem{DoIwMi01}
  \au{S.}{Doi}, \au{A.}{Iwatsuka}, \au{T.}{Mine}:
  \ti{The uniqueness of the integrated density of states for the Schr\"odinger operators
    with magnetic fields}
  \z{Math. Z.}{237}{335--371}{2001}
%
\bibitem{DoMa99} 
  \au{T.C.}{Dorlas}, \au{N.}{Macris}, \au{J.V.}{Pul{\'e}}:
  \ti{Characterization of the spectrum of the Landau Hamiltonian with
    delta impurities} 
  \z{Commun. Math. Phys.}{204}{367--396}{1999} 

\bibitem{DrKi86}
  \au{J.}{Droese}, \au{W.}{Kirsch}:
  \ti{The effect of boundary conditions on the density of states for
    random Schr\"odinger operators}
  \z{Stochastic Processes Appl.}{23}{169--175}{1986}
%


\bibitem{Fer75}  \def\infrench{[in French]}
  \au{X.M.}{Fernique}:
  \ti{Regularit{\'e} des trajectoires des fonctions al{\'e}atoires Gaussiennes}
  In:
  \bti{Ecole d'Et{\'e} de Probabilit{\'e}s de Saint-Flour IV - 1974}
  Hennequin, P.-L. (ed.), Lecture Notes in Mathematics {\bf 480}, 
  \pub[pp. 1--96 \infrench]{Springer}{Berlin}{1975} 



\bibitem{FiHu97b}
  \au{W.}{Fischer}, \au{T.}{Hupfer}, \au{H.}{Leschke}, \au{P.}{M{\"u}ller}: 
  \ti{Existence of the density of states for multi-dimensional
    continuum Schr{\"o}dinger operators with Gaussian random potentials}
  \z{Commun. Math. Phys.}{190}{133--141}{1997}


\bibitem{FiLeMu00}
  \au{W.}{Fischer}, \au{H.}{Leschke}, \au{P.}{M{\"u}ller}:
  \ti{Spectral localization by Gaussian random potentials in
    multi-dimensional continuous space}
  \z{J. Stat. Phys.}{101}{935--985}{2000}

\bibitem{Foc28}
        \au{V.}{Fock}:
        \ti{Bemerkung zur Quantelung des harmonischen Oszillators im Magnetfeld}
        \z[in German]{Z. Physik}{47}{ 446--448}{1928}

\bibitem{GT}
  \au{D.}{Gilbarg}, \au{N.S.}{Trudinger}:
  \bti{Elliptic partial differential equations of second order}
  $2^{\rm nd}$ edition,
  \pub{Springer}{Berlin}{1983}

\bibitem{GrRy}
  \au{I.S.}{Gradshteyn}, \au{I.M.}{Ryzhik}:
  \bti{Table of integrals, series, and products}
  Corrected and enlarged edition, 
  \pub{Academic}{San Diego}{1980}

\bibitem{HeScUh77}
  \au{H.}{Hess}, \au{R.}{Schrader}, \au{D.A.}{Uhlenbrock}:
  \ti{Domination of semigroups and generalization of Kato's inequality}
  \z{Duke Math. J.}{44}{893--904}{1977}

\bibitem{HLMW01}
  \au{T.}{Hupfer}, \au{H.}{Leschke}, \au{P.}{M\"uller}, \au{S.}{Warzel}:
  \ti{Existence and uniqueness of the integrated density of states for Schr\"odinger operators 
    with magnetic fields and unbounded random potentials}
  e-print math-ph/0010013 (2000)

\bibitem{HuLeWa00}
  \au{T.}{Hupfer}, \au{H.}{Leschke}, \au{S.}{Warzel}:
  \ti{Upper bounds on the density of states of single Landau levels broadened by Gaussian random potentials}
   e-print math-ph/0011010 (2000), to appear in \emph{J. Math. Phys.}

\bibitem{Kat72}
  \au{T.}{Kato}:
  \ti{Schr{\"o}dinger operators with singular potentials}
  \z{Israel J. Math.}{13}{135--148}{1972}

\bibitem{Kat78}
  \au{T.}{Kato}:  
  \ti{Remarks on Schr{\"o}dinger operators with vector potentials}
  \z{Integral Equations Oper. Theory}{1}{103--113}{1978} 

         
\bibitem{KatMas78}
  \au{T.}{Kato}, \au{K.}{Masuda}:
  \ti{Trotter's product formula for nonlinear semigroups generated 
      by the subdifferentials of convex functionals}
  \z{J. Math. Soc. Japan}{30}{169--178}{1978}


  
\bibitem{Kir89}
  \au{W.}{Kirsch}:        
  \ti{Random Schr{\"o}dinger operators: A course}
  In: 
  \bti{Schr{\"o}dinger operators}
  Holden, H., Jensen, A. (eds.),          
  Lecture Notes in Physics {\bf 345},    
  \pub[pp. 264--370]{Springer}{Berlin}{1989}

\bibitem{KirMar82a}
  \au{W.}{Kirsch}, \au{F.}{Martinelli}:
  \ti{On the ergodic properties of the spectrum of general random operators}
  \z{J. Reine Angew. Math.}{334}{141--156}{1982}

\bibitem{KirMar82}
  \au{W.}{Kirsch}, \au{F.}{Martinelli}:
  \ti{On the density of states of Schr{\"o}dinger operators 
    with a random potential}
  \z{J. Phys. A}{15}{2139--2156}{1982}

  
\bibitem{KuMeTi88}
  \au{I.V.}{Kukushkin}, \au{S.V.}{Meshkov}, \au{V.B.}{Timofeev}:
  \ti{Two-dimensional electron density of states in a transverse magnetic field}
  \z[Russian original: \z{Usp. Fiz. Nauk}{155}{219--264}{1988}]%
        {Sov. Phys. Usp.}{31}{511--534}{1988}

\bibitem{Lan30}
        \au{L.}{Landau}:
        \ti{Diamagnetismus der Metalle}
        \z[in German]{Z. Physik}{64}{629--637}{1930}

\bibitem{LiLo97}
        \au{E.H.}{Lieb}, \au{M.}{Loss}:
        \bti{Analysis}
        \pub{Am. Math. Soc.}{Providence, Rhode Island}{1997}

\bibitem{LiGr88}
  \au{I.M.}{Lifshits}, \au{S.A.}{Gredeskul}, \au{L.A.}{Pastur}:
  \bti{Introduction to the theory of disordered systems}  
  \pub*[Russian original:  \pub{Nauka}{Moscow}{1982}]%
  {Wiley}{New York}{1988}     

\bibitem{Lif95}
  \au{M.A.}{Lifshits}:
  \bti{Gaussian random functions}
  \pub{Kluwer}{Dordrecht}{1995}

\bibitem{LiMa97}
  \au{V.}{Liskevitch}, \au{A.}{Manavi}:
  \ti{Dominated semigroups with singular complex potentials}
  \z{J. Funct. Anal.}{151}{281--305}{1997}



\bibitem{Mat93}
  \au{H.}{Matsumoto}:
  \ti{On the integrated density of states for the Schr{\"o}dinger operators 
    with certain random electromagnetic potentials}
  \z{J. Math. Soc. Japan}{45}{197--214}{1993} 

\bibitem{MohRai94}
  \au{A.}{Mohamed}, \au{G.D.}{Raikov}:
  \ti{On the spectral theory of the Schr{\"o}dinger operator 
    with electromagnetic potential}
  In:
  \bti{Pseudo-differential calculus and mathematical physics}
  Demuth, M., Schrohe, E., Schulze, B.-W.(eds.),
  \pub[pp. 298--390]{Akademie}{Berlin}{1994}

\bibitem{Nak00}
  \au{S.}{Nakamura}:
  \ti{A remark on the Dirichlet-Neumann decoupling and the integrated
    density of states}
  \z{J. Funct. Anal.}{179}{136--152}{2001}

\bibitem{Nak77}
  \au{S.}{Nakao}:
  \ti{On the spectral distribution of the Schr{\"o}dinger operator with random potential}
  \z{Japan. J. Math.}{3}{111--139}{1977}

\bibitem{Pas71}
  \au{L.}{Pastur}:
  \ti{On the Schr{\"o}dinger equation with a random potential}
  \z[Russian original: \z{Teor. Mat. Fiz.}{6}{415--424}{1971}]{Theor. Math. Phys.}{6}{299--306}{1971}

\bibitem{Pas80}
  \au{L.}{Pastur}:
  \ti{Spectral properties of disordered systems in the 
    one-body approximation}
  \z{Commun. Math. Phys.}{75}{179--196}{1980}

\bibitem{PaFi92}
  \au{L.}{Pastur}, \au{A.}{Figotin}:
  \bti{Spectra of random and almost-periodic operators}
  \pub{Springer}{Berlin}{1992}


\bibitem{PeSe81} 
  \au{M.A.}{Perelmuter}, \au{Yu.A.}{Semenov}:
  \ti{On decoupling of finite singularities in the scattering theory for the 
        Schr{\"o}dinger operator with a magnetic field}
  \z{J. Math. Phys.}{22}{521--533}{1981}

 


\bibitem{ReSi80}
  \au{M.}{Reed}, \au{B.}{Simon}:
  \bti{Methods of modern mathematical physics I: Functional analysis} 
  Revised and enlarged edition,
  \pub{Academic}{San Diego}{1980}

\bibitem{ReSi75}
  \au{M.}{Reed}, \au{B.}{Simon}:
  \bti{Methods of modern mathematical physics II: Fourier analysis, 
    self-adjointness}
  \pub{Academic}{New York}{1975}

\bibitem{ReSi78}
  \au{M.}{Reed}, \au{B.}{Simon}:
  \bti{Methods of modern mathematical physics IV: Analysis of operators}
  \pub{Academic}{New York}{1978}


\bibitem{ShEf84}
  \au{B.I.}{Shklovskii}, \au{A.L.}{Efros}:
  \bti{Electronic properties of doped semiconductors}   
  \pub*[Russian original: \pub{Nauka}{Moscow}{1979}]%
     {Springer}{Berlin}{1984} 

\bibitem{Sim76}
  \au{B.}{Simon}:
  \ti{Universal diamagnetism of spinless Bose systems}
  \z{Phys. Rev. Lett.}{36}{1083--1084}{1976}
%
\bibitem{Sim77}
  \au{B.}{Simon}:
  \ti{An abstract Kato's inequality for generators of positivity preserving 
        semigroups}
  \z{Ind. Math. J.}{26}{1067--1073}{1977}

\bibitem{Sim79MM}
  \au{B.}{Simon}:
  \ti{Maximal and minimal Schr{\"o}dinger forms}
  \z{J. Operator Theory}{1}{37--47}{1979}

\bibitem{Sim79}
  \au{B.}{Simon}:
  \bti{Functional integration and quantum physics}
  \pub{Academic}{New York}{1979}


\bibitem{Sim79KI}
  \au{B.}{Simon}: 
  \ti{Kato's inequality and the comparison of semigroups}
  \z{J. Funct. Anal.}{32}{97--101}{1979} 



\bibitem{Sim00}
   \au{B.}{Simon}:
   \ti{Schr{\"o}dinger operators in the twenty-first century}
   In:
  \bti{Mathematical Physics 2000}
  Fokas, A., Grigoryan, A., Kibble, T., Zegarlinski, B. (eds.),
  \pub[pp. 283--288]{Imperial College Press}{London}{2000}

\bibitem{Sto01}
  \au{P.}{Stollmann}:
  \bti{Caught by disorder: Bound states in random media}
  \pub{Birkh\"auser}{Boston}{2001}
%
\bibitem{Uek94}
  \au{N.}{Ueki}:
  \ti{On spectra of random Schr{\"o}dinger operators with magnetic fields}
  \z{Osaka J. Math.}{31}{177--187}{1994}


\bibitem{Ves00}
  \au{I.}{Veseli\'c}:
  \ti{Wegner estimate for some indefinite Anderson-type Schr\"odinger operators}
  e-print mp\_arc 00-373 (2000)

\bibitem{Wan97} 
  \au{W.-M.}{Wang}:
  \ti{Microlocalization, percolation, and Anderson localization for
    the magnetic Schr{\"o}dinger operator with a random potential}
  \z{J. Funct. Anal.}{146}{1--26}{1997}


\bibitem{Weg81}
  \au{F.}{Wegner}:
  \ti{Bounds on the density of states in disordered systems}
  \z{Z. Phys. B}{44}{9--15}{1981}



\bibitem{Wey12}
\au{H.}{Weyl}:
\ti{Das asymptotische Verteilungsgesetz der Eigenwerte linearer partieller
Differentialgleichungen 
(mit einer Anwendung auf die Theorie der Hohlraumstrahlung)}
\z[in German]{Math. Ann.}{71}{441--479}{1912}

\bibitem{Zak64}
  \au{J.}{Zak}:
  \ti{Magnetic translation group}
  \z{Phys. Rev.}{134}{A1602--A1606}{1964}

\end{thebibliography}
\end{document}